\documentclass[12pt,]{article}
\usepackage{graphicx}
\usepackage{amssymb}

\input epsf

\newcommand{\D}{{\rm d}}

\newcommand{\BB}{\mbox{\boldmath$B$}}
\newcommand{\bb}{\mbox{\boldmath$b$}}

\newcommand{\CC}{\mbox{\boldmath$C$}}
\newcommand{\cc}{\mbox{\boldmath$c$}}
\newcommand{\DD}{\mbox{\boldmath$D$}}

\newcommand{\FF}{\mbox{\boldmath$F$}}
\newcommand{\ff}{\mbox{\boldmath$f$}}
\newcommand{\HH}{\mbox{\boldmath$H$}}
\newcommand{\bg}{\mbox{\boldmath$g$}}

\newcommand{\JJ}{\mbox{\boldmath$J$}}

\newcommand{\MM}{\mbox{\boldmath$M$}}
\newcommand{\NN}{\mbox{\boldmath$N$}}
\newcommand{\nn}{\mbox{\boldmath$n$}}
\newcommand{\uu}{\mbox{\boldmath$u$}}

\newcommand{\UU}{\mbox{\boldmath$U$}}

\newcommand{\vv}{\mbox{\boldmath$v$}}

\newcommand{\yy}{\mbox{\boldmath$y$}}
\newcommand{\xx}{\mbox{\boldmath$x$}}
\newcommand{\xxx}{\mbox{\boldmath$x$}}
\newcommand{\XX}{\mbox{\boldmath$X$}}

\newcommand{\bOmega}{\mbox{\boldmath$\Omega$}}
\newcommand{\bPi}{\mbox{\boldmath$\Pi$}}

\newcommand{\bDelta}{\mbox{\boldmath$\Delta$}}
\newcommand{\bgamma}{\mbox{\boldmath$\gamma$}}
\newcommand{\bmu}{\mbox{\boldmath$\mu$}}
\newcommand{\bnabla}{\mbox{\boldmath$\nabla$}}

\newcommand{\balpha}{\mbox{\boldmath$\alpha$}}
\newcommand{\bbeta}{\mbox{\boldmath$\beta$}}

\newcommand{\ww}{\mbox{\boldmath$w$}}

\newcommand{\xiprime}{\mbox{\boldmath${\xi}^{\prime}$}}

\newcommand{\bc}{\ensuremath{\mathbf{c}}}
\newcommand{\bu}{\ensuremath{\mathbf{u}}}
\newcommand{\bx}{\ensuremath{\mathbf{x}}}
\newcommand{\mxi}{\mbox{\boldmath${\xi}$}}

\begin{document}

\begin{center}
{\Large {\bf METHODS OF NONLINEAR KINETICS}}

{\it Alexander N. Gorban and Iliya V. Karlin}

ETH-Zentrum, Department of Materials, Institute of Polymers,

Sonneggstr. 3, ML J27, CH-8092 Z{\"u}rich, Switzerland

agorban@mat.ethz.ch, ikarlin@mat.ethz.ch

Institute of Computational Modeling SB RAS, Krasnoyarsk 660036, Russia

\end{center}

\tableofcontents

\clearpage

\section{The Boltzmann equation}

\subsection{The equation}

The {\it Boltzmann equation} is the first and the most celebrated nonlinear kinetic equation
introduced by the great Austrian scientist Ludwig Boltzmann in 1872 \cite{Boltzmann}. This equation
describes the dynamics of a moderately rarefied gas, taking into account the two processes: the free
flight of the particles, and their collisions. In its original version, the Boltzmann equation has
been formulated for particles represented by hard spheres. The physical condition of rarefaction
means that only pair collisions are taken into account, a mathematical specification of which is
given by the \index{Grad-Boltzmann limit}{\it Grad-Boltzmann limit} \cite{GradENC}: If $N$ is the
number of particles, and $\sigma$ is the diameter of the hard sphere, then the Boltzmann equation is
expected to hold when $N$ tends to infinity, $\sigma$ tends to zero, $N\sigma^3$ (the volume occupied
by the particles) tends to zero, while $N\sigma^2$ (the total collision cross section) remains
constant. The microscopic state of the gas at time $t$ is described by the one-body distribution
function $P(\xx,\vv,t)$, where $\xx$ is the position of the center of the particle, and $\vv$ is the
velocity of the particle. The distribution function is the probability density of finding the
particle at time $t$ within the infinitesimal phase space volume centered at the phase point
$(\xx,\vv)$. The collision mechanism of two hard spheres is presented by a relation between the
velocities of the particles before [$\vv$ and $\ww$] and after [$\vv'$ and $\ww'$] their impact:
\begin{eqnarray}
\vv'=\vv-\nn (\nn, \vv -\ww), \nonumber\\ \ww'=\ww+\nn (\nn, \vv -\ww),\nonumber
\end{eqnarray}
where $\nn$ is the unit vector along $\vv-\vv'$. Transformation of the velocities conserves the total
momentum of the pair of colliding particles $(\vv'+\ww'=\vv+\ww)$, and the total kinetic energy
$(\vv'^2+\ww'^2=\vv^2+\ww^2)$ The Boltzmann equation reads:
\begin{eqnarray}
{\partial P\over \partial t}+\left(\vv,{\partial P\over
\partial \xx
}\right)&=& \label{S1} N\sigma^2 \int_{R^3}\int_{B^-}(P(\xx,\vv',t)P(\xx,\ww',t) \nonumber \\ &&
-P(\xx,\vv,t)P(\xx,\ww,t))\mid(\ww-\vv,\nn)\mid  \,\D \ww  \,\D \nn,
\end{eqnarray}

\noindent where integration in $\ww$ is carried over the whole space $R^3$, while integration in
$\nn$ goes over a hemisphere $B^-=\{\nn \in S^2 \mid (\ww-\vv,\nn)<0 \}$ . This inequality
$(\ww-\vv,\nn)<0$ corresponds to the particles entering the collision. The nonlinear integral
operator in the right hand side of (\ref{S1}) is nonlocal in the velocity variable, and local in
space. The Boltzmann equation for arbitrary hard-core interaction is a generalization of the
Boltzmann equation for hard spheres under the proviso that the true infinite-range interaction
potential between the particles is cut-off at some distance. This generalization amounts to a
replacement,
\begin{eqnarray}
\sigma^2\mid(\ww-\vv,\nn)\mid \,\D\nn\to B(\theta,\mid\ww-\vv\mid)\,\D\theta \,\D\varepsilon,
\label{S2}
\end{eqnarray}
where function $B$ is determined by the interaction potential, and vector $\nn$ is identified with
two angles, $\theta$  and $\varepsilon$. In particular, for potentials proportional to the n-th
inverse power of the distance, the function $B$ reads,
\begin{eqnarray}
B(\theta,\mid\vv-\ww\mid)=\beta(\theta)\mid\vv-\ww\mid^{n-5\over n-1}.\label{S3}
\end{eqnarray}
In the special case $n=5$, function $B$ is independent of the magnitude of the relative velocity
(Maxwell molecules). \index{Maxwell molecules}Maxwell molecules occupy a distinct place in the theory
of the Boltzmann equation, they provide exact results. Three most important findings for the Maxwell
molecules are mentioned here: 1. The exact spectrum of the linearized Boltzmann collision integral,
found by Truesdell and Muncaster \cite{Truesdell}, 2. Exact transport coefficients found by Maxwell
even before the Boltzmann equation was formulated, 3. Exact solutions to the space-free model version
of the nonlinear Boltzmann equation. Pivotal results in this domain belong to Galkin \cite{Galkin}
who has found the general solution to the system of moment equations in a form of a series expansion,
to Bobylev, Krook and Wu \cite{Bob3,Krook,Krook1} who have found an exact solution of a particular
elegant closed form, and to Bobylev who has demonstrated the complete integrability of this dynamic
system \cite{BobEX}, the review of relaxation of spatially uniform dilute gases for several types of
interaction models, of exact solutions and related topics was given in \cite{ErnstPhysRep}.

A broad review of the Boltzmann equation and analysis of analytical solutions to kinetic models is
presented in the book of Cercignani \cite{Cercignani}. A modern account of rigorous results on the
Boltzmann equation is given in the book \cite{CeIlPu}. Proof of the existence theorem for the
Boltzmann equation was given by DiPerna and Lions \cite{DPL}.

It is customary to write the Boltzmann equation using another normalization of the distribution
function, $f(\xx,\vv,t) \,\D \xx  \,\D \vv$, taken in such a way that the function $f$ is compliant
with the definition of the \index{Hydrodynamic fields}hydrodynamic fields: the mass density
 $\rho$, the momentum density $\rho \uu$, and the energy density $\varepsilon$:
\begin{eqnarray}
\int f(\xx,\vv,t)m \,\D\vv &=&\rho(\xx,t),\nonumber \\ \int f(\xx,\vv,t)m \vv \, \D \vv
&=&\rho\uu(\xx,t), \label{S4}
\\ \int f(\xx,\vv,t)m{v^2\over 2} \,\D \vv &=&\varepsilon(\xx,t). \nonumber
\end{eqnarray}
Here $m$ is the particle's mass.

 The Boltzmann equation for the
distribution function $f$ reads,
\begin{eqnarray}
{\partial f\over \partial t}+\left(\vv,{\partial \over \partial \xx} f\right)=Q(f,f),\label{S5}
\end{eqnarray}
\index{Collision integral}where the nonlinear integral operator in the right hand side is the
Boltzmann collision integral,
\begin{eqnarray}
Q=\int_{R^3}\int_{B^-}(f(\vv')f(\ww')-f(\vv)f(\ww))B(\theta,\vv) \,\D \ww \,\D \theta  \,\D
\varepsilon. \label{S6}
\end{eqnarray}

Finally, we mention the following form of the Boltzmann collision integral (sometimes referred to as
the {\it scattering} or the {\it quasi-chemical} representation),
\begin{eqnarray}
Q=\int W(\vv,\ww\mid\vv',\ww')[(f(\vv')f(\ww')-f(\vv)f(\ww))] \,\D \ww  \,\D \ww'  \,\D \vv',
\label{S7}
\end{eqnarray}
where $W$ is a generalized function which is called the probability density of the elementary event,
\begin{eqnarray}
W=w(\vv,\ww \mid\vv', \ww')\delta(\vv+\ww-\vv' -\ww')\delta(v^2+w^2-v'^2-w'^2).\label{S8}
\end{eqnarray}

\subsection{The basic properties of the Boltzmann equation}

Generalized function W has the following symmetries:
\begin{eqnarray}
&&W(\vv', \ww'\mid\vv,\ww)\equiv W(\ww', \vv'\mid\vv,\ww) \nonumber \\ &&\equiv W(\vv',
\ww'\mid\ww,\vv)\equiv W(\vv, \ww\mid\vv',\ww').\label{S9}
\end{eqnarray}

The first two identities reflect the symmetry of the collision process with respect to labeling the
particles, whereas the last identity is the celebrated \index{Detailed balance}{\it detailed balance}
condition which is underpinned by the time-reversal symmetry of the microscopic (Newton's) equations
of motion. The basic properties of the Boltzmann equation are:

1. {\it Additive invariants of collision operator}:
\begin{eqnarray}
\int Q(f,f)\{1,\vv,v^2\} \,\D \vv=0, \label{S10}
\end{eqnarray}
for any function $f$, assuming integrals exist. Equality (\ref{S10}) reflects the fact that the
number of particles, the three components of particle's momentum, and the particle's energy are
conserved in collisions. Conservation laws (\ref{S10}) imply that the local hydrodynamic fields
(\ref{S4}) can change in time only due to redistribution over the space.

2. Zero point of the integral ($Q=0$) satisfy the equation (which is also called {\it the detailed
balance}): For almost all velocities,
\begin{eqnarray}
f(\vv',\xx,t)f(\ww',\xx,t)=f(\vv,\xx,t)f(\ww,\xx,t). \nonumber
\end{eqnarray}

3. Boltzmann's {\it local entropy production inequality}:\index{Entropy!production inequality}
\begin{eqnarray}
\sigma (\xx,t)=- k_{\rm B} \int  Q(f,f) \ln f  \,\D \vv \geq 0, \label{S11}
\end{eqnarray}
for any function $f$, assuming integrals exist. Dimensional \index{Boltzmann's constant}{\it
Boltzmann's constant} ($k_{\rm B}\approx 6\cdot 10^{-23}$Jole/Kelvin) in this expression serves for a
recalculation of the energy units into the absolute temperature units. Moreover, equality sign takes
place if $\ln f$ is a linear combination of the additive invariants of collision.

Distribution functions $f$ whose logarithm is a linear combination of additive collision invariants,
with coefficients dependent on $x$, are called \index{Maxwell distribution}{\it local Maxwell
distribution functions} $f_{\rm LM}$,
\begin{eqnarray}
f_{\rm LM}={\rho \over m} \left({2\pi k_{\rm B}T\over m}\right)^{-3/2}\exp\left({-m(\vv-\uu)^2\over
2k_{\rm B}T}\right).\label{S12}
\end{eqnarray}

\index{Local Maxwellians}Local Maxwellians are parametrized by values of five hydrodynamic variables,
$\rho$ , $\uu$ and $T$. This parametrization is consistent with the definitions of the hydrodynamic
fields (\ref{S4}), $\int f_{\rm LM}\{m,m\vv,mv^2/2\} \,\D \vv =(\rho, \rho\uu,\varepsilon)$ provided
the relation between the energy and the kinetic temperature $T$, holds, $\varepsilon ={3\rho \over 2
m k_{\rm B}T}$.

4. Boltzmann's {\it {\boldmath $H$} theorem}: The function
\begin{eqnarray}
S[f]=-k_{\rm B}\int f \ln f  \,\D \vv, \label{S13}
\end{eqnarray}
is called the \index{Entropy!density}{\it entropy density}\footnote{From physical point of view a
value of the function $f$ can be treated as dimensional quantity, but if one changes the scale and
multiplies $f$ by a positive number $\nu$ then $S[f]$ transforms into $\nu S[f]+\nu \ln{\nu}\int f
\,\D \vv$. For a closed system the corresponding transformation of the entropy is an inhomogeneous
linear transformation with constant coefficients.}. The {\it local {\boldmath $H$} theorem} for
distribution functions independent of space states that the rate of the entropy density increase is
equal to the nonnegative entropy production,
\begin{eqnarray}
{ \,\D S\over  \,\D t}=\sigma\geq0. \label{S14}
\end{eqnarray}

Thus, if no space dependence is concerned, the Boltzmann equation describes relaxation to the unique
global Maxwellian (whose parameters are fixed by initial conditions), and the entropy density grows
monotonically along the solutions. Mathematical specifications of this property has been initialized
by Carleman, and many estimations of the entropy growth were obtained over the past two decades. In
the case of space-dependent distribution functions, the local entropy density obeys the {\it entropy
balance equation}:
\begin{eqnarray}
{\partial S(\xx,t)\over \partial t}+\left({\partial \over
\partial \xx},\JJ_s(\xx,t)\right)=\sigma (\xx,t)\geq0,\label{S15}
\end{eqnarray}

\noindent where $\JJ_s$  is the entropy flux, $\JJ_s(\xx,t)=-k_{\rm B} \int \ln f(\xx,t)\vv f(\xx,t)
\,\D \vv.$ For suitable boundary conditions, such as, specularly reflecting or at the infinity, the
entropy flux gives no contribution to the equation for the {\it total entropy}, $S_{tot}=\int
S(\xx,t) \,\D \xx$
  and its rate of changes is then equal to the nonnegative total entropy
production $\sigma_{tot}=\int \sigma(\xx,t) \,\D \xx$ \index{Global {\boldmath $H$} theorem}(the {\it
global {\boldmath $H$} theorem)}. For more general boundary conditions which maintain the entropy
influx the global $H$ theorem needs to be modified. A detailed discussion of this question is given
by Cercignani \cite{Cercignani}. The local Maxwellian is also specified as the maximizer of the
Boltzmann entropy function (\ref{S13}), subject to fixed hydrodynamic constraints (\ref{S4}). For
this reason, the local Maxwellian is also termed as the local equilibrium distribution function.

\subsection{Linearized collision integral}

Linearization of the Boltzmann integral around the local equilibrium results in the linear integral
operator,
\begin{eqnarray}
&&Lh(\vv)=\int W(\vv,\ww \mid\vv',\ww') f_{\rm LM}(\vv)f_{\rm LM}(\ww) \nonumber\\ && \;\; \times
\left[{h(\vv')\over f_{\rm LM}(\vv')}+{h(\ww')\over f_{\rm LM}(\ww')}-{h(\vv)\over f_{\rm
LM}(\vv)}-{h(\ww)\over f_{\rm LM}(\ww)}\right] \,\D \ww'  \,\D \vv'  \,\D \ww.
\end{eqnarray}
{\it Linearized collision integral} is symmetric with respect to \index{Entropic!scalar
product}scalar product defined by the second derivative of the entropy functional,
\begin{eqnarray}
\int f_{\rm LM}^{-1}(\vv)g(\vv)L h(\vv) \,\D  \vv = \int f_{\rm LM}^{-1}(\vv)h(\vv)Lg(\vv) \,\D
\vv,\nonumber
\end{eqnarray}
it is nonpositively definite,
\begin{eqnarray}
\int f_{\rm LM}^{-1}(\vv)h(\vv)L h(\vv) \,\D \vv\leq 0,\nonumber
\end{eqnarray}
where equality sign takes place if the function $hf_{\rm LM}^{-1}$ is a linear combination of
collision invariants, which characterize the null-space of the operator $L$. Spectrum of the
linearized collision integral is well studied in the case of the small angle cut-off.

\section{Phenomenology and Quasi-chemical representation of the Boltzmann equation}

\index{Quasi-chemical representation}Boltzmann's original derivation of his collision integral was
based on a phenomenological ``bookkeeping" of the gain and of the loss of probability density in the
collision process. This derivation postulates that the rate of gain $G$ equals
\begin{eqnarray}
G=\int W^+(\vv,\ww\mid\vv',\ww')f(\vv')f(\ww') \,\D \vv'  \,\D \ww'  \,\D \ww, \nonumber
\end{eqnarray}
while the rate of loss is
\begin{eqnarray}
L=\int W^-(\vv,\ww\mid\vv',\ww')f(\vv)f(\ww) \,\D \vv'  \,\D \ww'  \,\D \ww.\nonumber
\end{eqnarray}

The form of the gain and of the loss, containing products of one-body distribution functions in place
of the two-body distribution, constitutes the famous Stosszahlansatz. The Boltzmann collision
integral follows now as ($G-L$), subject to the detailed balance for the rates of individual
collisions,
\begin{eqnarray}
W^+(\vv,\ww\mid\vv',\ww')=W^-(\vv,\ww\mid\vv',\ww').\nonumber
\end{eqnarray}

This representation for interactions different from hard spheres requires also the cut-off of
functions $\beta$  (\ref{S3}) at small angles. The gain-loss form of the collision integral makes it
evident that the detailed balance for the rates of individual collisions is sufficient to prove the
local $H$ theorem. A weaker condition which is also sufficient to establish the $H$ theorem was first
derived by Stueckelberg \cite{Stueck} (so-called \index{Semi-detailed balance}{\it  semi-detailed
balance}), and later generalized {\it to inequalities of concordance}\index{Inequalities of
concordance} \cite{G1}:
\begin{eqnarray}
\int  \,\D \vv' \int  \,\D \ww'(W^+(\vv,\ww \mid \vv',\ww')-W^-(\vv,\ww \mid \vv',\ww'))\geq
0,\nonumber\\ \int  \,\D \vv \int  \,\D \ww (W^+(\vv,\ww \mid \vv',\ww')-W^-(\vv,\ww \mid
\vv',\ww'))\leq 0.\nonumber
\end{eqnarray}

The semi-detailed balance follows from these expressions if the inequality signes are replaced by
equalities.

The pattern of Boltzmann's phenomenological approach is often used in order to construct nonlinear
kinetic models. In particular, nonlinear {\it equations of chemical kinetics} are based on this idea:
If $n$ chemical species $A_i$  participate in a complex chemical reaction,
\begin{eqnarray}
\sum_i \alpha_{si}A_i\leftrightarrow \sum_i \beta_{si}A_i, \nonumber
\end{eqnarray}
where $\alpha_{si}$  and $\beta_{si}$  are nonnegative integers ({\it stoichiometric coefficients})
then equations of chemical kinetics for the concentrations of species $c_j$  are written
\begin{eqnarray}
{ \,\D c_i\over  \,\D t}=\sum_{s=1}^n (\beta_{si}-\alpha_{si})\left[\varphi_s^+\exp
\left(\sum_{j=1}^n{\partial G\over \partial c_j}\alpha_{sj}\right) -\varphi_s^-\exp
\left(\sum_{j=1}^n{\partial G\over \partial c_j}\beta_{sj}\right)\right].\nonumber
\end{eqnarray}

Functions $\varphi_s^+$  and $\varphi_s^-$  are interpreted as constants of the direct and of the
inverse reactions, while the function $G$ is an analog of the Boltzmann's $H$-function.

Modern derivation of the Boltzmann equation, initialized by the seminal work of N.N. Bogoliubov
\cite{Bogol}, seek a replacement condition for the Stosszahlansatz which would be more closely
related to many-particle dynamics. Different conditions has been formulated by D.N. Zubarev
\cite{Zubarev}, R.M. Lewis \cite{Lew} and others. The advantage of these formulations is the
possibility to systematically find corrections not included in the Stosszahlansatz.

\section{Kinetic models}

Mathematical complications caused by the nonlinearly Boltzmann collision integral are traced back to
the Stosszahlansatz. Several approaches were developed in order to simplify the Boltzmann equation.
Such simplifications are termed kinetic models. Various kinetic models preserve certain features of
the Boltzmann equation, while sacrificing the rest of them. The most well known kinetic model which
preserves the $H$ theorem is the nonlinear \index{BGK model}Bhatnagar-Gross-Krook model (BGK)
\cite{BGK}. The BGK collision integral reads:
\begin{eqnarray}\label{BGKsource}
Q_{\rm BGK}=-{1\over \tau}(f-f_{\rm LM}(f)).\nonumber
\end{eqnarray}
The time parameter $\tau > 0$ is interpreted as a characteristic relaxation time to the local
Maxwellian. The BGK collision integral is the nonlinear operator: Parameters of the local Maxwellian
are identified with the values of the corresponding moments of the distribution function $f$. This
nonlinearly is of ``lower dimension" than in the Boltzmann collision integral because $f_{\rm LM}(f)$
is a nonlinear function of only the moments of $f$ whereas the Boltzmann collision integral is
nonlinear in the distribution function $f$ itself. This type of simplification introduced by the BGK
approach is closely related to the family of so-called mean-field approximations in statistical
mechanics. By its construction, the BGK collision integral preserves the following three properties
of the Boltzmann equation: additive invariants of collision, uniqueness of the equilibrium, and the
$H$ theorem. A class of kinetic models which generalized the BGK model to quasiequilibrium
approximations of a general form is described as follows: The quasiequilibrium  $f^*$ for the set of
linear functionales $M(f)$ is a distribution function $f^*(M)(\xx,\vv)$  which maximizes the entropy
under fixed values of functions $M$. The \index{Quasiequilibrium!models}Quasiequilibrium (QE) models
are characterized by the collision integral \cite{GKMod},
\begin{eqnarray}
Q_{\rm QE}(f)=-{1\over \tau}[f-f^*(M(f))]+Q(f^*(M(f)),f^*(M(f))).\nonumber
\end{eqnarray}
Same as in the case of the BGK collision integral, operator $Q_{\rm QE}$ is nonlinear in the moments
$M$ only. The QE models preserve the following properties of the Boltzmann collision operator:
additive invariants, uniqueness of the equilibrium, and the $H$ theorem, provided the relaxation time
$\tau$ to the quasiequilibrium is sufficiently small \cite{GKMod}. A different nonlinear model was
proposed by Lebowitz, Frisch and Helfand \cite{LFHMod}:
\begin{eqnarray}
Q_D=D\left({\partial \over \partial \vv}{\partial \over \partial \vv }f+{m\over k_{\rm B}T}{\partial
\over
\partial \vv}(\vv -\uu(f))f\right).\nonumber
\end{eqnarray}
The collision integral has the form of the self-consistent Fokker-Planck operator, describing
diffusion (in the velocity space) in the self-consistent potential. Diffusion coefficient $D>0$ may
depend on the distribution function $f$. Operator $Q_D$ preserves the same properties of the
Boltzmann collision operator as the BGK model. The kinetic BGK model has been used for obtaining
exact solutions of gasdynamic problems, especially its linearized form for stationary problems.
Linearized BGK collision model has been extended to model more precisely the linearized Boltzmann
collision integral \cite{Cercignani}.

\section{Methods of reduced description}

One of the major issues raised by the Boltzmann equation is the problem of the reduced description.
Equations of hydrodynamics constitute a closed set of equations for the hydrodynamic field (local
density, local momentum, and local temperature). From the standpoint of the Boltzmann equation, these
quantities are low-order moments of the one-body distribution function, or, in other words, the
macroscopic variables. The problem of the reduced description consists in giving an answer to the
following two questions:

1. What are the conditions under which the macroscopic description sets in?

2. How to derive equations for the macroscopic variables from kinetic equations?

The classical methods of reduced description for the Boltzmann equation are: the Hilbert method, the
Chapman-Enskog method, and the Grad moment method.

\subsection{The Hilbert method}

In 1911, David Hilbert introduced the notion of \index{Normal solution}normal solutions, $$f_{\rm
H}(\vv,\,n(\xx,t),\,\uu(\xx,t),\,T(\xx,t)),$$

\noindent that is, solutions to the Boltzmann equation which depend on space and time only through
five hydrodynamic fields \cite{Hilbert}.

The normal solutions are found from a singularly perturbed Boltzmann equation,
\begin{eqnarray}
D_tf={1\over \varepsilon}Q(f,f),\label{S16}
\end{eqnarray}
where $\varepsilon$ is a small parameter, and $$D_t f\equiv {\partial \over \partial
t}f+\left(\vv,{\partial \over \partial \xx}\right)f.$$ Physically, parameter $\varepsilon$
corresponds to the \index{Knudsen number}Knudsen number, the ratio between the mean free path of the
molecules between collisions, and the characteristic scale of variation of the hydrodynamic fields.
In the Hilbert method, one seeks functions $n(\xx,t),\,\uu(\xx,t),\,T(\xx,t)$, such that the normal
solution in the form of the Hilbert expansion,
\begin{eqnarray}
f_{\rm H}=\sum_{i=0}^{\infty}\varepsilon^i f_{\rm H}^{(i)}\label{S17}
\end{eqnarray}
satisfies the (\ref{S16}) order by order. Hilbert was able to demonstrate that this is formally
possible. Substituting (\ref{S17}) into (\ref{S16}), and matching various order in $\varepsilon$, we
obtain the sequence of integral equations
\begin{eqnarray}
&&Q(f_{\rm H}^{(0)},f_{\rm H}^{(0)})=0,\label{S18}\\ &&Lf_{\rm H}^{(1)}=D_tf_{\rm
H}^{(0)},\label{S19}\\ &&Lf_{\rm H}^{(2)}=D_tf_{\rm H}^{(1)}-2Q(f_{\rm H}^{(0)},f_{\rm
H}^{(1)}),\label{S20}
\end{eqnarray}
and so on for higher orders. Here $L$ is the linearized collision integral. From (\ref{S18}), it
follows that $f_{\rm H}^{(0)}$  is the local Maxwellian with parameters not yet determined. The
Fredholm alternative, as applied to the second equation from (\ref{S19}) results in

a) Solvability condition,
\begin{eqnarray}
\int D_tf_{\rm H}^{(0)}\{1,\vv,v^2\} \,\D \vv =0,\nonumber
\end{eqnarray}
which is the set of the compressible Euler equations of the non-viscous hydrodynamics. Solution to
the Euler equation determine the parameters of the Maxwellian  $f_{\rm H}^0$.

b) General solution $f_{\rm H}^{(1)}=f_{\rm H}^{(1)1}+f_{\rm H}^{(1)2}$, where $f_{\rm H}^{(1)1}$ is
the special solution to the linear integral equation (\ref{S19}), and $f_{\rm H}^{(1)2}$ is yet
undetermined linear combination of the additive invariants of collision.

c) Solvability condition to the next equation (\ref{S19}) determines coefficients of the function
$f_{\rm H}^{(1)2}$ in terms of solutions to linear hyperbolic differential equations,
\begin{eqnarray}
\int D_t(f_{\rm H}^{(1)1}+f_{\rm H}^{(1)2})\{1,\vv,v^2\} \,\D \vv=0.\nonumber
\end{eqnarray}
Hilbert was able to demonstrate that this procedure of constructing the normal solution can be
carried out to arbitrary order $n$, where the function $f_{\rm H}^{(n)}$ is determined from the
solvability condition at the next, $(n+1)$-th order. In order to summarize, implementation of the
Hilbert method requires solutions for the functions $n(\xx,t),\,\uu(\xx,t)$, and $T(\xx,t)$  obtained
from a sequence of partial differential equations.

\subsection{The Chapman-Enskog method}

A completely different approach to the reduced description was invented in 1917 by David Enskog
\cite{Ens}, and independently by Sidney Chapman \cite{Chapman}. The key innovation was to seek an
expansion of the time derivatives of the hydrodynamic variables rather than seeking the time-space
dependencies of these functions, as in the Hilbert method.

The Chapman-Enskog method starts also with the singularly perturbed Boltzmann equation, and with the
expansion
\begin{eqnarray}
f_{\rm CE}=\sum_{n=0}^{\infty} \varepsilon^n f_{\rm CE}^{(n)}.\nonumber
\end{eqnarray}
However, the procedure of evaluation of the functions $f_{\rm CE}^{(n)}$ differs from the Hilbert
method:
\begin{eqnarray}
Q(f_{\rm CE}^{(0)},f_{\rm CE}^{(0)})&=&0,\label{S21}\\ Lf_{\rm CE}^{(1)}&=&-Q(f_{\rm CE}^{(0)},f_{\rm
CE}^{(0)})+{\partial^{(0)}\over
\partial t }f_{\rm CE}^{(0)}+\left(\vv, {\partial \over \partial
\xx}\right)f_{\rm CE}^{(0)}.\label{S22}
\end{eqnarray}
Operator $\partial^{(0)} /\partial t$  is defined from the expansion of the right hand side of
hydrodynamic equation,
\begin{eqnarray}
{\partial^{(0)}\over \partial t}\{\rho,\rho\uu,e\}\equiv -\int \left\{m,m\vv,{mv^2\over 2}\right\}
\left( \vv,{\partial \over \partial \xx} \right) f^{(0)}_{\rm CE} \,\D \vv.\label{S23}
\end{eqnarray}
From (\ref{S21}), function $f_{\rm CE}^{(0)}$ is again the local Maxwellian, whereas (\ref{S23}) is
the Euler equations, and $\partial^{(0)} /\partial t$ acts on various functions $g(\rho,\rho\uu,e)$
according to the chain rule,
\begin{eqnarray}
{\partial^{(0)}\over
\partial t }g={\partial g\over \partial
\rho }{\partial^{(0)} \over \partial t}\rho +{\partial g\over
\partial (\rho \uu)}{\partial^{(0)} \over \partial t}(\rho \uu) +
{\partial g\over \partial e}{\partial^{(0)}  \over \partial t}e,\nonumber
\end{eqnarray}
while the time derivatives ${\partial^{(0)} \over \partial t}$ of the hydrodynamic fields are
expressed using the right hand side of (\ref{S23}).

 The result of the Chapman-Enskog definition of the
time derivative ${\partial^{(0)} \over \partial t}$, is that the Fredholm alternative is satisfied by
the right hand side of (\ref{S22}). Finally, the solution to the homogeneous equation is set to be
zero by the requirement that the hydrodynamic variables as defined by the function
$f^{(0)}+\varepsilon f^{(1)}$  coincide with the parameters of the local Maxwellian  $f^{(0)}$:
\begin{eqnarray}
\int \{1,\vv,v^2\}f_{\rm CE}^{(1)} \,\D \vv=0.\nonumber
\end{eqnarray}

The first correction $f_{\rm CE}^{(1)}$ of the Chapman-Enskog method adds the terms
\begin{eqnarray}
{\partial^{(1)}\over \partial t }\{\rho,\rho\uu,e\}=-\int \left\{m,m\vv,{mv^2\over
2}\right\}\left(\vv,{\partial \over
\partial \xx}\right)f_{\rm CE}^{(1)} \,\D \vv \nonumber
\end{eqnarray}
to the time derivatives of the hydrodynamic fields. These terms correspond to the dissipative
hydrodynamics where viscous momentum transfer and heat transfer are in the Navier-Stokes and Fourier
form. The Chapman-Enskog method was the first true success of the Boltzmann equation since it made it
possible to derive macroscopic equation without a priori guessing (the generalization of the
Boltzmann equation onto mixtures predicted existence of the thermodiffusion before it has been found
experimentally), and to express transport coefficients in terms of microscopic particle's
interaction.

However, higher-order corrections of the Chapman-Enskog method, resulting in hydrodynamic equations
with higher derivatives (Burnett hydrodynamic equations) face severe difficulties both from the
theoretical, as well as from the practical sides. In particular, they result in unphysical
instabilities of the equilibrium.

\subsection{The Grad moment method}

In 1949, Harold Grad extended the basic assumption of the Hilbert and the Chapman-Enskog methods (the
space and time dependence of normal solutions is mediated by the five hydrodynamic moments)
\cite{Grad}. A physical rationale behind the Grad moment method is an assumption of the decomposition
of motions:

\noindent(i). During the time of order $\tau$, a set of distinguished moments $M'$ (which include the
hydrodynamic moments and a subset of higher-order moments) does not change significantly as compared
to the rest of the moments $M''$ (the fast dynamics).

\noindent(ii). Towards the end of the fast evolution, the values of the moments $M''$ become
unambiguously determined by the values of the distinguished moments $M'$.

\noindent(iii). On the time of order $\theta \gg \tau$, dynamics of the distribution function is
determined by the dynamics of the distinguished moments while the rest of the moments remain to be
determined by the distinguished moments (the slow evolution period).

Implementation of this picture requires an ansatz for the distribution function in order to represent
the set of states visited in the course of the slow evolution. In Grad's method, these representative
sets are finite-order truncations of an expansion of the distribution functions in terms of Hermite
velocity tensors:
\begin{eqnarray}
f_G(M',\vv)=f_{\rm LM}(\rho,\uu,E,\vv)[1+\sum_{(\alpha)}^N a_{(\alpha)}(M')H_{(\alpha)}(\vv-\uu)]
,\label{S24}
\end{eqnarray}
 where $H_{(\alpha)}(\vv-\uu)$ are Hermite tensor polynomials,
orthogonal with the weight $f_{\rm LM}$, while coefficient $a_{(\alpha)}(M')$ are known functions of
the distinguished moments $M'$, and $N$ is the highest order of $M'$. Other moments are functions of
$M'$: $M''=M''(f_G(M'))$.

Slow evolution of distinguished moments is found upon substitution of (\ref{S24}) into the Boltzmann
equation and finding the moments of the resulting expression ({\it Grad's moment equations}).
Following Grad, this extremely simple approximation can be improved by extending the list of
distinguished moments. The most well known is Grad's thirteen-moment approximation where the set of
distinguished moments consists of five hydrodynamic moments, five components of the traceless stress
tensor $\sigma_{ij}=\int m[(v_i-u_i)(v_j-u_j)-\delta_{ij}(\vv-\uu)^2/3]f \,\D \vv,$ and of the three
components of the heat flux vector $q_i=\int (v_i-u_i)m(\vv-\uu)^2/2 f \,\D \vv$.

 The decomposition of motions hypothesis
cannot be evaluated for its validity within the framework of Grad's approach. It is not surprising
therefore that Grad's methods failed to work in situations where it was (unmotivatedly) supposed to,
primarily, in the phenomena with sharp time-space dependence such as the strong shock wave. On the
other hand, Grad's method was quite successful for describing transition between parabolic and
hyperbolic propagation, in particular, the second sound effect in massive solids at low temperatures,
and, in general, situations slightly deviating from the classical Navier-Stokes-Fourier domain.
Finally, the Grad method has been important background for development of phenomenological
nonequilibrium thermodynamics based on hyperbolic first-order equation, the so-called EIT (extended
irreversible thermodynamics \cite{EIT,Mueller}).

\subsection{Special approximations}

Special approximations of the solutions to the Boltzmann equation were found for several problems,
and which perform better than results of ``regular" procedures. The most well known is the ansatz
introduced independently by Mott-Smith and Tamm for the strong shock wave problem: The (stationary)
distribution function is thought as
\begin{eqnarray}
f_{\rm TMS}(a(x))=(1-a(x))f_{+}+a(x)f_{-},\label{S25}
\end{eqnarray}
where $f_{\pm}$ are upstream and downstream Maxwell distribution functions, whereas $a(x)$ is an
undetermined scalar function of the coordinate along the shock tube.

Equation for function $a(x)$ has to be found upon substitution of (\ref{S25}) into the Bolltzmann
equation, and upon integration with some velocity-dependent function  $\varphi(\vv)$. Two general
problems arise with the special approximation thus constructed: Which function $\varphi(\vv)$ should
be taken, and how to find correction to the ansatz like  (\ref{S25})?

\subsection{The method of invariant manifold}

The general problem of reduced description for dissipative system was recognized as the problem of
finding stable invariant manifolds in the space of distribution functions
\cite{CMIM,GKAMSE92,GK1,GKTTSP94}. The notion of invariant manifold generalizes the normal solution
in the Hilbert and in the Chapman-Enskog method, and the finite-moment sets of distribution function
in the Grad method: If $\Omega$ is a smooth manifold in the space of distribution function, and if
$f_{\Omega}$ is an element of $\Omega$, then $\Omega$ is invariant with respect to the dynamic
system,
\begin{eqnarray}
&&{\D  f\over  \D  t}=J(f), \label{S26}\\&& \mbox{if}  \: J(f_{\Omega})\in T_{f_{\Omega}}\Omega,  \:
\mbox{for all} \: f_{\Omega}\in \Omega, \label{S27}
\end{eqnarray}
where $T_{f_{\Omega}}\Omega$ is the tangent space of the manifold  $\Omega$ at the point
$f_{\Omega}$. Application of the invariant manifold idea to dissipative systems is based on
iterations, progressively improving the initial approximation, and it involves the following steps:
construction of the thermodynamic projector and iterations for the invariance condition

\subsubsection{Thermodynamic projector}

 Given a manifold $\Omega$  (not obligatory invariant), the macroscopic
dynamics on this manifold is defined by the {\it macroscopic vector field}, which is the result of a
projection of vectors $J(f_{\Omega})$ onto the tangent bundle $T\Omega$. The thermodynamic projector
$ P^*_{f_{\Omega}}$ takes advantage of dissipativity:
\begin{eqnarray}
\mbox{ker} P^*_{f_{\Omega}}\subseteq \mbox{ker}D_fS\mid_{f_{\Omega}},\label{S28}
\end{eqnarray}

\noindent where $D_fS\mid_{f_{\Omega}}$ is the differential of the entropy evaluated in $f_{\Omega}$.

\index{Thermodynamicity condition}This condition of thermodynamicity means that each state of the
manifold $\Omega$ is regarded as the result of decomposition of motions occurring near $\Omega$: The
state $f_{\Omega}$ is the maximum entropy state on the set of states $f_{\Omega}+\mbox{ker}
P^*_{f_{\Omega}}$. Condition of thermodynamicity does not define projector completely; rather, it is
the condition that should be satisfied by any projector used to define the macroscopic vector field,
$J'_{\Omega}=P^*_{f_{\Omega}}J(f_{\Omega})$. For, once the condition (\ref{S28}) is met, the
macroscopic vector field preserves dissipativity of the original microscopic vector field $J(f)$:
\begin{eqnarray}
D_fS\mid_{f_{\Omega}}\cdot P^*_{f_{\Omega}}(J(f_{\Omega}))\geq 0\mbox{ for all
}\,f_{\Omega}\in\Omega.\nonumber
\end{eqnarray}

The thermodynamic projector is the formalization of the assumption that $\Omega$ is the manifold of
slow motion: If a fast relaxation takes place at least in a neighborhood of $\Omega$, then the states
visited in this process before arriving at $f_{\Omega}$ belong to $\mbox{ker} P^*_{f_{\Omega}}$. In
general, $P^*_{f_{\Omega}}$ depends in a non-trivial way on $f_{\Omega}$.

\subsubsection{Iterations for the invariance condition}

 The invariance
condition for the manifold $\Omega$ reads,
\begin{eqnarray}
P_{\Omega}(J(f_{\Omega}))-J(f_{\Omega})=0,\nonumber
\end{eqnarray}
here $P_{\Omega}$ is arbitrary (not obligatory thermodynamic) projector onto the tangent bundle of
$\Omega$. The invariance condition is considered as an equation which is solved iteratively, starting
with initial approximation $\Omega_0$. On the $(n+1)-$th iteration, the correction
$f^{(n+1)}=f^{(n)}+\delta f^{(n+1)}$ is found from linear equations,
\begin{eqnarray}
D_f J^*_n\delta f^{(n+1)}&=&P^*_{n}J(f^{(n)})-J(f^{(n)}),\nonumber
\\ P^*_{n}\delta f^{(n+1)}&=&0,\label{S29}
\end{eqnarray}
here $D_f J^*_n$ is the linear self-adjoint operator with respect to the scalar product by the second
differential of the entropy $D^2_f S\mid_{f^{(n)}}$.

Together with the above-mentioned principle of thermodynamic projecting, the {\it self-adjoint
linearization} implements the assumption about the decomposition of motions around the $n$'th
approximation. The \index{Self-adjoint linearization}self-adjoint linearization of the Boltzmann
collision integral $Q$ (\ref{S7}) around a distribution function $f$ is given by the formula,
\begin{eqnarray}
D_f Q^{\rm SYM}\delta f&=& \int W(\vv,\ww, \mid \vv',\ww'){f(\vv)f(\ww)+f(\vv')f(\ww')\over 2}
\nonumber
\\ &\times&\left[{\delta f (\vv')\over f(\vv')}+{\delta f(\ww')\over f(\ww')}-{\delta f (\vv)\over
f(\vv)}-{\delta f(\ww)\over f(\ww)}\right] \,\D \ww'  \,\D \vv'  \,\D \ww. \label{S30}
\end{eqnarray}

 If $f=f_{\rm LM}$, the self-adjoint operator (\ref{S30}) becomes the
 linearized collision integral.

The method of invariant manifold is the iterative process: $$(f^{(n)},P^*_{n})\rightarrow
(f^{(n+1)},P^*_{n})\rightarrow (f^{(n+1)},P^*_{n+1})$$ On the each 1-st step of the iteration, the
linear equation (\ref{S29}) is solved with the projector known from the previous iteration. On the
each 2-nd step, the projector is updated, following the thermodynamic construction.

The method of invariant manifold can be further simplified if smallness parameters are known.

The proliferation of the procedure in comparison to the Chapman-Enskog method is essentially twofold:

First, the projector is made dependent on the manifold. This enlarges the set of admissible
approximations.

Second, the method is based on iteration rather than a series expansion in a smallness parameter.
Importance of iteration procedures is well understood in physics, in particular, in the
renormalization group approach to reducing the description in equilibrium statistical mechanics, and
in the Kolmogorov- Arnold-Moser theory of finite-dimensional Hamiltonian systems.

\subsection{Quasiequilibrium approximations}

Important generalization of the Grad moment method is the concept of the {\it quasiequilibrium
approximations} already mentioned above (we discuss this approximation in detail in a separate
section). The quasiequilibrium distribution function for a set of distinguished moment $M=m(f)$
maximizes the entropy density $S$ for fixed $M$. The quasiequilibrium manifold $\Omega^*(M)$ is the
collection of the quasiequilibrium distribution functions for all admissible values of $M$. The
quasiequilibrium approximation is the simplest and extremely useful (not only in the kinetic theory
itself) implementation of the hypothesis about a decomposition of motions: If $M$ are considered as
slow variables, then states which could be visited in the course of rapid motion in the vicinity of
$\Omega^*(M)$ belong to the planes $$\Gamma_{M}=\{f\mid m(f-f^*(M))=0\}.$$ In that respect, the
thermodynamic construction in the method of invariant manifold is a generalization of the
quasiequilibrium approximation where the given manifold is equipped with a quasiequilibrium structure
by choosing appropriately the macroscopic variables of the slow motion. In contrast to the
quasiequilibrium, the macroscopic variables thus constructed are not obligatory moments. A textbook
example of the quasiequilibrium approximation is the generalized Gaussian function for $M=\{\rho,
\rho \uu,P\}$ where $P_{ij}=\int v_iv_j f  \,\D \vv$ is the pressure tensor.

The thermodynamic projector $P^*$ for a quasiequilibrium approximation was first introduced by B.
Robertson \cite{Robertson} (in a different context of conservative dynamics and for a special case of
the Gibbs-Shannon entropy). It acts on a function $\Psi$ as follows $$P^*_{M}\Psi =\sum_i{\partial
f^*\over
\partial M_i}\int m_i \Psi  \,\D \vv,$$ where $M=\int m_i f  \,\D \vv$. The quasiequilibrium approximation
does not exist if the highest order moment is an odd-order polynomial of velocity (therefore, there
exists no quasiequilibrium for thirteen Grad's moments), and a regularization is then required.
Otherwise, the Grad moment approximation is the first-order expansion of the quasiequilibrium around
the local Maxwellian.

\section{Discrete velocity models}

If the number of microscopic velocities is reduced drastically to only a finite set, the resulting
discrete velocity models, continuous in time and in space, can still mimic the gas-dynamic flows.
This idea was introduced in Broadwell's paper in 1963 to mimic the strong shock wave
\cite{Broadwell}.

Further important development of this idea was due to Cabannes and Gatignol in the seventies who
introduced a systematic class of discrete velocity models \cite{Discret2}. The structure of the
collision operators in the discrete velocity models mimics the polynomial character of the Boltzmann
collision integral. Discrete velocity models are implemented numerically by using the natural
operator splitting in which each update due to free flight is followed by the collision update, the
idea which dates back to Grad. One of the most important recent results is the proof of convergence
of the discrete velocity models with pair collisions to the Boltzmann collision integral \cite{Palc}.

\section{Direct simulation}

Besides the analytical approach, direct numerical simulation of Boltzmann-type nonlinear kinetic
equations have been developed since mid of 1960's, beginning with the seminal works of Bird
\cite{GBird,Oran}. The basis of the approach is a representation of the Boltzmann gas by a set of
particles whose dynamics is modeled as a sequence of free propagation and collisions. The modeling of
collisions uses a random choice of pairs of particles inside the cells of the space, and changing the
velocities of these pairs in such a way as to comply with the conservation laws, and in accordance
with the kernel of the Boltzmann collision integral. At present, there exists a variety of this
scheme known under the common title of the Direct Simulation Monte-Carlo method \cite{GBird,Oran}.
The DSMC, in particular, provides data to test various analytical theories.

\section{Lattice Gas and Lattice Boltzmann models}

 Since the mid of 1980's, the kinetic-theory based approach to simulation of complex
macroscopic phenomena such as hydrodynamics has been developed. The main idea of the approach is the
construction of minimal kinetic system in such a way that their long-time and large-scale limit
matches the desired macroscopic equations. For this purpose, the fully discrete (in time, space, and
velocity) nonlinear kinetic equations are considered on sufficiently isotropic lattices, where the
links represent the discrete velocities of fictitious particles. In the earlier version of the
lattice methods, the particle-based picture has been exploited, subject to the exclusion rule (one or
zero particle per lattice link) [the lattice gas model \cite{Lgas} ]. Most of the present versions
use the distribution function picture, where populations of the links are non-integer [the lattice
Boltzmann model \cite{Mcnamara,LB1,LB2,Higuera,Benzi}]. Discrete-time dynamics consists of a
propagation step where populations are transmitted to adjacent links and collision step where
populations of the links at each node of the lattice are equilibrated by a certain rule. Most of the
present versions use the BGK-type equilibration, where the local equilibrium is constructed in such a
way as to match desired macroscopic equations. The lattice Boltzmann method is a useful approach for
computational fluid dynamics, effectively compliant with parallel architectures. The proof of the $H$
theorem for the Lattice gas models is based on the semi-detailed (or Stueckelberg's) balance
principle. The proof of the $H$ theorem in the framework of the lattice Boltzmann method has been
only very recently achieved \cite{LB3,KGSBPRL,KFOeEPL,AKPRE00,AKJSP,AKOeEPL}.

\section{Minimal Boltzmann models for flows at low Knudsen number}
\label{ELB}

In this section, we present  a new discrete Boltzmann model which has the correct thermodynamics, and
which  reproduces Navier-Stokes-Fourier equation in the  hydrodynamic limit. Discrete velocities are
taken as zeros of the Hermite  polynomials, and the maximum entropy Grad moment method, together with
the Gauss-Hermite quadrature is used in order to derive  the discrete $H$ function. Corresponding
local equilibria are found analytically. The  numerical implementation involves a novel
discretization scheme consistent with the $H$ theorem. Discretization scheme is extended to a
derivation of the diffusive boundary  condition. Some numerical results of the simulation of the
model  for different flow problems are presented.

Kinetic theory based simulation schemes, in the context of the computational fluid dynamics, have
attracted considerable attention in the last decade.  In particular, the lattice Boltzmann method has
emerged as an efficient alternative to the traditional Navier-Stokes solvers \cite{LB1,LB2}. In the
spirit of the  kinetic theory of gases, in this method simple pseudo-particle kinetics in introduced
in such a way that hydrodynamic equations are obtained as its large-scale long-time limit. The method
is essentially a particular discrete Boltzmann model with a simplified collision mechanism.  The
emphasis of this model is on recovering desired macroscopic dynamics with minimal computational
costs. However, it turns out that a vital feature of the kinetic theory, the  $H$ theorem, was lost
in this simplification process \cite{LB1,LB2,LB3}. This deficiency in the method had hindered the
further development of the method as a tool for simulating thermal hydrodynamics \cite{LB3}.

The aim of this work is to describe a recently proposed method \cite{AKOeEPL},  which allows
construction of lattice Boltzmann like methods from the considerations of the classical kinetic
theory \cite{Cercignani,GKMod}. The resulting scheme is computationally as efficient as the standard
lattice Boltzmann method. The  scheme is based on preserving continuous (in time) as well as the
discrete $H$ theorem \cite{Cercignani,LB3}, and hereby it  ensures the non-linear stability of the
resulting numerical algorithm.

\subsection{Lattice Boltzmann Method}

We shall first describe the standard lattice Boltzmann model. The basic setup for the method is as
follows:
 Let $f_i( \bx,t)$ be a population
of discrete velocities $\bc_i$, $i=1,\dots,b$, at position $\bx$ and time $t$. Hydrodynamic fields
are the first few  moments of populations, namely
\begin{equation}
\sum_{i=1}^b f_i \{ 1,\  c_{i\alpha},\ c_i^2 \}=\{\rho,\ \rho u_{\alpha},\ \rho DT+ \rho u^2 \},
\end{equation}
where $\rho$ is the mass density of the fluid, $\rho u_{\alpha}$ is the momentum density, and $\rho D
T+ \rho u^2$  is the energy density. In our notation, $\alpha=1,\dots, D$, denotes the spatial
directions, where $D$ is the spatial dimension. In the case of athermal hydrodynamics, the list of
independent hydrodynamic fields  consists of only the  mass and the momentum density. Typically it is
assumed that during the collision, the populations relax to their equilibrium value with  a single
relaxation time  $\tau$ (Bhatnagar-Gross-Krook (BGK) form of the collision \cite{Cercignani}) . The
kinetic equation under this assumption reads,
\begin{equation}
\label{LBM}
\partial_t f_i+ c_{i\alpha} \partial_{\alpha}f_i =
-\tau^{-1} \left(f_i- f_i^{\rm eq}  \right),
\end{equation}
 where the form of the
 local equilibrium population, $f_i^{\rm eq}$  remains  to be specified.

 The set of discrete velocities is usually chosen to  form links of a
 Bravais lattice (with possibly several sub-lattices). This
choice of discrete velocity  facilitates in an efficient discretization using the method of
 characteristics. The condition that the local
 hydrodynamic variables  are invariants of the  collision,
\begin{equation}
\label{mom} \sum_{i=1}^b f^{\rm eq}_i \{ 1,\  c_{i\alpha},\ c_i^2 \}=\{\rho,\ \rho u_{\alpha},\ \rho
DT+ \rho u^2 \},
\end{equation}
imposes a set of restrictions
 on the form of the  equilibrium population .
(Note that, in the athermal case the energy constraint on the local equilibrium is  excluded from
this list.)  Chapman-Enskog analysis of  (\ref{LBM}), with unknown equilibrium satisfying
  (\ref{mom}), shows  that in order to  recover the desired
hydrodynamic equations, in
 the large-scale long-time limit certain higher order moments of the
 equilibrium, which are involved in  the  Chapman-Enskog
analysis,  need to be  in the same form as in the classical kinetic theory.
 Further, it is assumed that the equilibrium population
 can be expressed as a polynomial in the hydrodynamic fields.
These  approximation allow for an explicit construction of the lattice
 Boltzmann method with correct hydrodynamics \cite{LB2}.

The above mentioned  way of constructing  kinetic models is not unique for constructioning the
equilibrium or for selection of the set of discrete velocities. It has been long known that not every
choice of the discrete velocities and the corresponding equilibria results in a stable simulation
algorithm.
 In the case of the athermal hydrodynamics,
the set of the discrete velocities and corresponding polynomials for the equilibrium have been found
which lead to a relatively stable realization. However, the problem of stability becomes particularly
severe when thermal modes are included. The reason for this failure is attributed to the lack of the
$H$ theorem for the method. Recently, it was shown how the method should be equipped with the $H$
theorem in the isothermal case.  The  numerical stability of the method is considerably enhanced by
the incorporation of the $H$ theorem  \cite{AKPRE00,AKJSP}.

\subsection{Entropic discrete BGK model}

The key of the construction is the $H$ function which needs to be found for the dynamics under
consideration \cite{AKJSP,KFOeEPL}. By definition,  the local equilibrium  is then the  minimizer of
the corresponding $H$ function \cite{LB3}, under constraints of the local conservation laws
(\ref{mom}). Furthermore, if  the corresponding local equilibria are known analytically, we can use
the BGK model. However, an explicit knowledge of the equilibrium is not a fundamental requirement for
constructing lattice Boltzmann like method. In other cases, when only the
 $H$ function is known analytically, an alternative single
 relaxation time model can be used, which circumvents the problem of
 finding the local equilibria in a closed form \cite{AKMod}.

The key question is how to find the correct set of discrete velocities and the corresponding $H$
function. In order to answer this, we remind the reader of the important observation on the relation
between the discretization of the velocity space and the well known Grad's moment method
\cite{Cercignani,DLBM}. Namely, if discrete velocities are constructed from  zeros of the Hermite
polynomials, the method of discrete velocity is essentially equivalent to  Grad's  moment method
based on the expansion of the distribution function around a fixed Maxwellian distribution function.
This link ensures that the resulting hydrodynamics is the correct one. However, expanding the local
equilibria means giving up the advantage of having the $H$ theorem \cite{DLBM}. This problem is
circumvented, if we directly evaluate the Boltzmann $H$ function for the discrete case and construct
the dynamics using this $H$ function. The idea is to use the Gauss-Hermite quadrature in order to
construct the set of discrete velocities and the $H$ function. This is essentially equivalent to
using  the entropic Grad's method (the maximum entropy approximation) \cite{GKMod}. In the next
section we describe how the method works.

\subsubsection{Discrete $H$ function}

 Boltzmann's $H$ function written in terms of the one-particle
distribution function $F({\xx}, \bc)$ is $H=\int F\ln F \,\D \bc$, where $\bc$ is the continuous
velocity. Close to the local equilibrium, this integral is naturally approximated  using  the
Gauss-Hermite quadrature. A direct evaluation of Boltzmann's $H$ function by a quadrature reads,
\begin{equation}
\label{app:H} H_{\{w_i,\bc_i\}}=\sum_{i=1}^{b} f_{i}\ln\left(\frac{f_{i}}{w_i} \right).
\end{equation}
Here $w_i$ is the weight associated with the $i$th discrete velocity $\bc_i$, and the particles mass
and the Boltzmann constant $k_{\rm B}$ are set equal to the unity. Discrete discrete function
$f_i({\xx})$ are related to values of the continuous distribution function at the nodes of the
quadrature as
 $f_i({\xx})=w_i(2\, \pi \, T_0)^{(D/2)}
\exp(c^2_i/(2\,T_0))F({\xx},
 {\cc}_i)$.
The discrete-velocity entropy functions (\ref{app:H}) for various $\{w_i,\bc_i\}$ is the unique input
for all our constructions below.
 In this setting, the  set of discrete velocities  corresponds to
zeroes of the Hermite polynomials in $\bc/ \sqrt{2 \, T_0}$. In the next section, it will be shown
how the order of the Hermite polynomial, whose zeroes constitute the set of discrete velocities,
affect the  dynamics.

\subsection{Hydrodynamics}

As the order of the Hermite polynomial is increased (this  will correspond to increasing number of
discrete velocities), the discrete $H_{\{w_i,\bc_i\}}$ function becomes a better  approximation to
the continuous Boltzmann $H$ function. Thus, with the increase in the order of the Hermite polynomial
a better approximation to the hydrodynamics is constructed. This behavior is demonstrated in Table
\ref{Tab: DiscV}.

\begin{table}[t]
 \caption{\label{Tab: DiscV}Reconstruction of the macroscopic dynamics with the increase
 in the order of the Hermite polynomial.}
\begin{tabular}{l|l|l|l|l}
\hline {Order of} &  {Independent} & {Discrete}  &{Weights}  & {Target equation}
 \\
{Polynomial} & {variables} & {velocities in $1$-$D$} & & \\
 \hline
$2$ & $\rho$ & $\pm 1$ & $\frac{1}{2}$ &Diffusion\\\hline
$3$ & $\rho$, $\rho u_{\alpha}$ &$0$,  $\pm \,\sqrt{3 \,T_0  }$  & $\frac{2}{3}$,$\,\frac{1}{6}$
&Athermal Navier-Stokes, $O(u^2)$\\\hline
$4$ & $\rho$, $\rho u_{\alpha}$, $E$ & $\pm \,a$, $\,\pm \,b$  & $\frac{T_0}{4 a^2}$, $\,\frac{T_0}{4
b^2}$& Thermal Navier-Stokes, $O(\theta^2)$ \\
    &                                            &   &
           &Athermal Navier-Stokes, $O(u^3)$
\\\hline
\end{tabular}
\end{table}

where $a=\sqrt{3-\sqrt{6}}(\,T_0)^{1/2}$, and $b=\sqrt{3+\sqrt{6}}(\,T_0)^{1/2}$.
  In higher dimensions, the discrete
velocities are  products of the discrete velocities in one dimension,  and the weights are
constructed by multiplying the weights associated with the each component direction.

   In order to give a better prospective of the method, we shall
   discuss a few examples of the Table \ref{Tab: DiscV} in details.

 \subsubsection{Athermal hydrodynamics}

If the discrete velocities are formed using  roots of the third-order Hermite polynomials (see row
$1$ of  table \ref{Tab: DiscV}) the Navier-Stokes equation is reproduced up to the order ${O}(u^2)$.
The explicit expression for the equilibrium distribution, obtained from solving the minimization
problem, reads:
 \begin{eqnarray}
\label{TED} && f^{\rm eq}_i= \rho w_i\times \\ && \prod_{\alpha=1}^{D} \left[ \left(2 -\sqrt{1+
(u_\alpha/c_{\rm s})^2}\right) \left(\frac{(2/\sqrt{3})(u_\alpha/c_{\rm s})+ \sqrt{1+
(u_\alpha/c_{\rm s})^2}}{1-u_\alpha/(\sqrt{3}c_{\rm s})} \right)^{c_{i\alpha}/(\sqrt{3}c_{\rm
s})}\right],  \nonumber
\end{eqnarray}
where $c_{\rm s}=\sqrt{\,T_0}$ is speed of sound. Note that the exponent,
$(c_{i\alpha}/(\sqrt{3}c_{\rm s}))$, in (\ref{TED}) takes
 the values $\pm 1, \, \mbox{and}\,\,  0$ only and the resulting
 expressions for the equilibrium can be simplified in each dimensions.
Equilibria (\ref{TED}) are positive definite for $u_{\alpha}<\sqrt{3}c_{\rm s}$. Without going into
details of derivation, we mention that the equilibrium (\ref{TED}) is the product of $D$
one-dimensional solutions which were found earlier in the paper \cite{AKPRE00} via a direct
minimization procedure, and it coincides with the classical equilibrium of the one dimensional
Broadwell model \cite{BroadwellO}. Also, the $H$ function for the two dimensional case was found
earlier  as the solution  to functional equation which imposes the requirement of having correct
stress tensor \cite{KFOeEPL}. The factorization over spatial dimensions is pertinent to the
derivation, and is quite similar to the familiar  property of Maxwell's distribution function. As the
higher order moments are not enforced, we need to check the behavior of the higher order moments.
Relevant higher-order moments of the equilibrium distribution, needed to establish the athermal
hydrodynamics in the framework of  Chapman-Enskog method are the equilibrium pressure tensor,
$P_{\alpha \beta}^{\rm eq}=\sum_{i}f_i^{\rm eq}c_{i\alpha}c_{i\beta}$,
 the equilibrium third-order moments,
$Q_{\alpha \beta \gamma }^{\rm eq}=\sum_{i}f_i^{\rm
  eq}c_{i\alpha}c_{i\beta}c_{i\gamma}$, and  the equilibrium fourth order moment
 $R_{\alpha\beta}^{\rm eq}=\sum_i c_{i\,\alpha}c_{i\,\beta} c^2f^{\rm
   eq}_i$. For the case of the athermal hydrodynamics, to recover
 Navier-Stokes equation only the
 equilibrium pressure tensor and  the equilibrium third-order moments
 are required to be in the correct form.  The deviation of these higher
order moments, as compared to the continuous case is reported in table \ref{Tab: Error}.

\begin{table}[t]
\caption{\label{Tab: Error} Behavior of higher order moments, in comparison to the
  continuous case. The symbol $\Delta$ denotes the difference from the
  continuous case.}
\begin{tabular}{c | c |  c | c}\hline
&{ $\Delta \, P_{\alpha \beta}^{\rm eq}$}
 &  {$\Delta \, Q_{\alpha
  \beta \gamma}^{\rm eq}$}
  & {$\Delta \,R_{\alpha \beta}^{\rm eq} $}
\\ \hline
{Athermal case} & $O(u^4)$ & $O(u^3)$& \\\hline {Thermal case*} & $O(u^8)$ & $O(u\theta^2)$,
 $O(u^3 \theta)$, and $O(u^5)$& $O(\theta^2)$, $O(u^2 \theta^2)$, and
$O(u^4)$\\ \hline
\end{tabular}
{*\scriptsize{$\theta = (T_0- T)/T_0$ is the deviation of the temperature from the reference value.}}
\end{table}

We see that the present equilibrium results in the correct pressure tensor  only to the order
${O}(u^4)$. However, this  is sufficient for simulations of low Mach number flows.  For example, for
$D=2$, even at a high  velocity, $u_\alpha = 0.25\sqrt{\,T_0}$, the error in the  pressure is less
than $2\%$.

 As  the derivation of the equilibrium (\ref{TED}) is based on the entropy
function (\ref{app:H}) which, in turn, is produced by the same quadrature as the polynomial
equilibria of the lattice Boltzmann method \cite{DLBM}, it is not surprising that expansions of the
function to the order ${O}(u^2)$ coincides with the polynomial equilibria (for $D=2$) of the standard
lattice Boltzmann method found earlier.

\subsubsection{Thermal hydrodynamics}

In precisely the same way, the minimal entropic kinetic model for the thermal case requires zeroes of
fourth-order  Hermite polynomials.
 In order to evaluate Lagrange multipliers in the
formal solution to the minimization problem, $$f^{\rm eq}_i=w_i \exp{\left( A +  \BB \cdot \bc_i + C
\,c_i^2\right)},$$ we first make an important observation that they can be computed exactly for
$\bu=0$ and {\it any} temperature $T$ within the positivity interval, $a^2<T<b^2$:
\begin{eqnarray}
B_\alpha = 0, \quad C_0 =\frac{1}{(b^2 - a^2)} \log { \left( \frac{w_a\, (T-a^2)}{w_b\, ( b^2 -
T)}\right)}, \nonumber \\ A_0 =  \log{\left( \frac{\rho\, (b^2 -T)^D }{ (2 w_a)^D ( b^2
-a^2)^D}\right)} - D\,a^2 C_0.
\end{eqnarray}
With this, we find the equilibrium  at zero average velocity and arbitrary temperature,
\begin{equation}
\label{TH0} f^{\rm eq}_i = \frac{\rho\, w_i}{2^D(b^2 - a^2)^D}\times \prod_{\alpha=1}^{D}
\left(\frac{b^2 - T  }{w_a} \right)^{ \left(\frac{b^2 - c_{i \alpha}^2 }{b^2 - a^2} \right)}
\left(\frac{T - a^2 }{w_b} \right)^{ \left(\frac{c_{i \alpha}^2 - a^2 }{b^2 - a^2} \right)} .
\end{equation}
Factorization over spatial components is clearly seen in this solution. Once the exact solution for
zero velocity  is known, extension to $\bu\ne0$ is easily found by perturbation. The first few terms
of the expanded Lagrange multipliers are:
\begin{eqnarray*}
A &=& A_0-\frac{T}{( T - a^2)(b^2 - T)}u^2+ O(u^4),
\\
B_{\alpha}&=&\frac{u_\alpha}{ T}  + \frac{(T - T_{0})^2}{2D T^4} \left(D\, u_\beta u_\theta u_\gamma
\delta_{\alpha \beta \gamma \theta} -3u^2\,u_\alpha \right) + O(u^5),
\\
C&=&C_0+ \frac{ a^2( b^2 -T) -T (b^2 -3T)  }{2DT^2( T - a^2)(b^2 - T) }u^2+O(u^4).
\end{eqnarray*}
For the actual numerical implementation, the equilibrium distribution function can be calculated
analytically, up to any order of accuracy required, by this procedure. Accuracy of  the relevant
higher order moments in this case is shown
 in the Table \ref{Tab: Error}. Once the error in these terms are small, the
present
 model reconstructs   the full thermal hydrodynamic equations.

While in the athermal case the closeness of the resulting macroscopic equations to the Navier-Stokes
equations is controlled solely by the deviations from zero of the average velocity, in the thermal
regime an additional control is due to variations of the temperature away from the reference value.
This means that not only  the actual velocity should be much less than the heat velocity
 but also the fractional  temperature deviation from $T_0$
 should be small, $| T- T_0|/T_0\ll 1$.  However, by increasing the
reference temperature, one gets a wider operating window of the present model. Another important
remark is about the use of the this thermal model for the athermal Navier-Stokes equation. If the
temperature is fixed at the reference value $T=T_0$, the pressure tensor and the third moment $Q^{\rm
eq}_{\alpha\beta\gamma}$ becomes exact  to the order ${O}(u^5)$, unlike in
 the  second-order accurate standard lattice  Boltzmann models and
 the athermal model constructed above.

\subsubsection{Transport coefficients}

When the single relaxation time BGK model (\ref{LBM})  is used with the present thermal  equilibrium,
the resulting transport coefficients are as follows: For $D=1$, the  kinematic viscosity $\nu$ is
equal to zero, while the thermal conductivity $\kappa$ is $\kappa = (3/2)(\tau \,\rho)T$. For $D>1$,
we have $\nu =(\tau \rho) T$,  and $ \kappa=((D+2)/2)(\tau \, \rho)T$.  The equality of  Prandtl
number, $c_p \nu\kappa^{-1}$, to unity
 is the well-known limitation of the single relaxation time model which
is cured by using multi-relaxation time models.  The density dependence of the transport coefficient
in the BGK model can be circumvented by  renormalizing  the relaxation time as $\tau'=\tau \, \rho$.

\subsection{Thermodynamics}

In our   construction of the  discrete velocity model, the main focus is on achieving a good
approximation of the Boltzmann $H$ function. Thus, we can expect that the correct thermodynamics will
be preserved (within the accuracy of the discretization), even in the discrete case. Indeed,   the
local equilibrium entropy, $S = - k_{\rm B} H_{\{w_i, \bc_i\}}(f^{\rm eq})$, for the thermal model
satisfies the usual expression for the entropy  of the ideal monoatomic gas to the overall order of
approximation of the method,
\begin{equation}
 S= \rho\, k_{\rm B} \ln{\left(T^{D/2}/ \rho\right) } + O(u^4,
\theta^2).
 \end{equation}

\subsection{Boundary condition}

Once we have developed a systematic way of obtaining the discrete velocity model from the continuous
case, the same idea can be extended easily to derive the  boundary condition.
 The methodology for obtaining the boundary
  condition is same as that for obtaining the discrete $H$ function. We
  shall start with the boundary condition written in the integral form
(see \cite{Cercignani})
  and then use the \index{Gauss-Hermite quadrature}Gauss-Hermite quadrature to obtain the boundary
  condition for the discrete case. However,  the  integral appearing in the boundary condition
is over half space. At this point, we should use  a counterpart of Hermite polynomial defined on the
half space only.
 However, once we have chosen the quadrature points for the interior (for evaluating the $H$
  function), we no longer have freedom to choose the quadrature points
  independently for the evaluation of the boundary condition. For that
 reason,  we are
  forced to use the same quadrature in this case too and then we need
  to evaluate the error in each case considered.  It turns out that
 for the case of the \index{Diffusive wall}diffusive wall in the athermal case, this way of
 evaluation of the half-space integral of the boundary condition
 is of the  accuracy  ${O}(u^2)$, as in the bulk. For the present
 purpose, a wall $\partial R$ is completely specified at any point $(
{\xx} \in\partial R )$ by the knowledge
 of the inward unit normal
 ${\nn}$, the  wall
 temperature $T_{\rm w }$ and the wall velocity
${\UU}_{\rm w }$.  The explicit expression for the
 boundary condition is
\begin{equation}
\label{DBC}
 f_i
 =  \frac{
 \sum_{\mbox{\boldmath {\scriptsize $\xi^{\prime}$}}_i \cdot {\nn} < 0  }
|(\xiprime_i\cdot{\nn})| f_{i^{\prime}} }{ \sum_{\mbox{\boldmath {\scriptsize $\xi^{\prime}$}}_i
\cdot {\nn} < 0  } |(\xiprime_i\cdot{\nn})| f_{i^{\prime}}^{\rm eq}( U_{\rm w }, \rho_{\rm w })}
f^{\rm eq}_i( U_{\rm w },  \rho_{\rm w }) ,   \qquad  ( \left( {\cc}_i -{\UU}_{\rm w }\right)\cdot
{\nn} > 0),
\end{equation}
  Here,  $\mxi$ denotes the molecular velocity  in a frame of reference
moving
 with the wall velocity and is equals to
$  {\cc} - {\UU}_{\rm w }$.

\subsection{Spatial and time  discretization}

In this subsection, we derive a discretization scheme for the discrete
 Boltzmann equation (\ref{LBM}), which leads to the lattice
 Boltzmann equation.

To begin with, we note that  the left hand side of  (\ref{LBM})   denotes the convection process, in
which no dissipation is generated (thus no entropy production). On the other hand, the  right hand
side of the equation denotes the generation of the entropy due to the
 relaxation of the population  to its equilibrium value (a local
 event). In order to explore this physical picture further, we look at
 the  solution of the (\ref{LBM}) after time $\delta t$
\begin{equation}
f_i(\delta t) = \exp{( - \delta t \left ( \bc_i \cdot \bnabla + L \right) )} f_i(0) +
{O}\left(\left(\frac{\delta t}{\tau}\right)^2\right),
\end{equation}
where, $L$ denotes the collision operator and in the case of the BGK
  approximation considered here is given as $L f_i =
  \tau^{-1} (f^{\rm eq}_i -f_i)$.
After performing an expansion in powers of commutator of the
  collision and the convection, we can
write
\begin{equation}
f_i(\delta t) = \exp{( -\delta t L) } \, \left[ \exp{( -\delta t \left ( \bc_i \cdot \bnabla \right)
)} f_i(0) \right] + {O}\left(\left(\frac{\delta t}{\tau}\right)^2\right),
\end{equation}
This, expansion has reduced, the problem into two analytically solvable problems   of the relaxation
free convection and of a local  relaxation process.  The solution reads,
\begin{eqnarray}
\label{exactLB} f_i(\bx, \delta  t) &=& f_i( \bx - \bc_i \delta  t , 0)  \\ & +&\left(1-
  \exp{\left[-\frac{\delta t}{\tau}\right]} \right)
\left( f_i^{\rm eq} - f_i( \bx - \bc_i \delta  t , 0)\right) + {O}\left(\left(\frac{\delta
t}{\tau}\right)^2\right). \nonumber
\end{eqnarray}

 The athermal lattice Boltzmann method utilizes this solution in a
 efficient manner by   tuning  the time step and the grid spacing in
such a way
 that $\bx - \bc_i \delta  t $ is always a grid point. This solves,
 the convection process  in a trivial fashion.
   In the thermal  case, a  modification in the algorithm is
 needed to minimize the mismatch of the spatial grid with the grid in
 the velocity space (set of discrete velocities). The aim is chose the
 $\delta t $ and $\delta \bx $ in such a way that mismatch is reduced
 to minimum. Presently, we are working to achieve such a
 discretization.

However, a  important modification in (\ref{exactLB}) is needed if we want to achieve the zero
viscosity limit. In the zero viscosity limit ($\tau \rightarrow 0$)  (\ref{exactLB}) will require
$\delta t \rightarrow 0$.    This problem can be avoided if many short collision
  steps can be lumped together in some ways. This can be done by modifying the
         discrete kinetic equation (\ref{exactLB}) as
\begin{equation}
\label{lbe} f_i(\bx, \delta  t) = f_i( \bx - \bc_i \delta  t , 0)  + 2 \beta \delta t \left( f_i^{\rm
eq} - f_i( \bx - \bc_i \delta  t , 0)\right),
\end{equation}
where the  discrete inverse relaxation time is $\beta =1/ \left({2 \tau}+  \delta  t \right)$. In the
limit of  $\tau \gg \delta t$, (\ref{exactLB}) and the modified  (\ref{lbe}) are identical. However,
in the limit of $\tau$ tending to zero, these two equation have two different relaxational behavior.
Note, that only short time dynamics is changed here to achieve rapid convergence towards
hydrodynamics. Here we remind that the relation
\begin{equation}
 1-
  \exp{\left[-\frac{\delta t}{\tau}\right]} =
\frac{\frac{\delta  t}{\tau}}{1 + \frac{\delta t}{2 \tau}} + {O}\left(\left(\frac{\delta
t}{\tau}\right)^3 \right).
\end{equation}
is a Pad\'e  approximation rather than a Taylor series expansion. This also explains, why in the
standard isothermal lattice Boltzmann method the viscosity is $\nu =  (1/\beta - \delta t) c_s^2/2$.

The discrete equation (\ref{lbe}) is successfully used as for various
 isothermal flow simulations \cite{LB1,LB2}. However,
a drawback of this over-relaxation scheme is the loss of $H$ theorem. In (\ref{lbe}), the positivity
of the population is not guaranteed. This may leads to numerical instability  in many cases (where
during the simulation the populations might drift  far away from the equilibrium)
\cite{AKJSP,AKMod,AK4}.

\subsubsection{A geometric procedure for restoring the discrete $H$ theorem}

\begin{figure}[t]
\centering{
\includegraphics[height=.25\textheight]{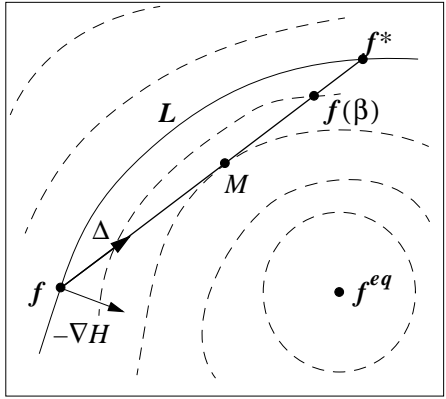}
\caption{\label{LBFig1} Stabilization procedure. Curves represent entropy levels, surrounding the local equilibrium
$\ff^{\rm eq}$. The solid curve $L$ is the entropy level with the value $H(\ff)=H(\ff^*)$, where
$\ff$ is the initial, and $\ff^*$ is the auxiliary population. The vector $\bDelta$ represents the
collision integral, the sharp angle between $\bDelta$ and the vector $-\bnabla H$ reflects the
entropy production inequality. The point $\MM$ is the solution to (\ref{MDDLB}). The result of the
collision update is represented by the point $\ff(\beta)$. The choice of $\beta$ shown corresponds to
the `over-relaxation': $H(\ff(\beta))>H(\MM)$ but $H(\ff(\beta))<H(\ff)$. The particular case of the
BGK collision (not shown) would be represented by a vector $\bDelta_{\rm BGK}$, pointing from $\ff$
towards $\ff^{\rm eq}$, in which case $\MM=\ff^{\rm eq}$. }}
\end{figure}

The advantage of the over-relaxation in achieving large time steps can be retained, and the  problem
of the numerical instability can be removed  if populations are over-relaxed in such a way that $H$
function does not increase in this process \cite{AKJSP,KFOeEPL}. To do so, we modify (\ref{lbe}) as
\begin{equation}
\label{Hlbe} f_i(\bx, \delta  t) = f_i( \bx - \bc_i \delta  t , 0)  + \alpha \beta \left( f_i^{\rm
eq} - f_i( \bx - \bc_i \delta  t , 0)\right),
\end{equation}
where, $\alpha$ is the solution to the equation,
\begin{equation}
\label{MDDLB} H(\ff) =H\left(\ff^* \right).
\end{equation}
here $\ff$ denotes $b$ dimensional vector consisting of all
    populations and $\ff^*= \ff - \alpha
\left(\ff- \ff^{\rm eq}  \right)$.
     First, the distribution is
    over-relaxed along the path dictated by local collision (with
    $\beta =1$) to a point
    of  equal entropy. Afterwards, a second collision with the
    relaxation parameter $\beta$  ensures that in the hydrodynamic
    limit correct viscosity coefficient is recovered.  The algorithm
    is illustrated  graphically in the Fig. \ref{LBFig1} .
      This way of implementing the $H$ theorem in the method
    ensures the non-linear stability.
  For the BGK model it can be shown that close to local equilibrium
  $\alpha = 2 \delta t$. The details of the implementation of the discrete $H$
    theorem is discussed in the paper \cite{AKJSP}.

\subsection{Numerical Experiments}

\begin{figure}[t]
\centering{
\includegraphics[scale=0.35]{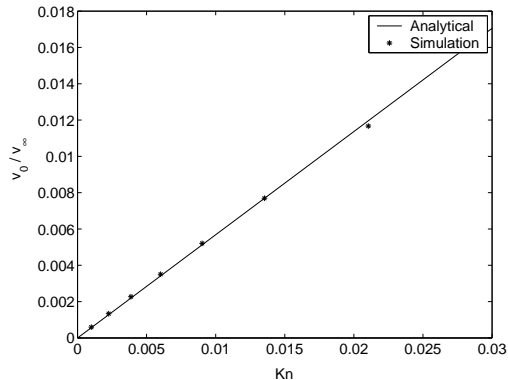}
\caption{ \label{LBFig2}  Relative slip observed at the wall in the simulation of the
  Kramers' problem for
shear rate   $a = 0.001 $,  box length $L=32$, $v_\infty = a \times L = 0.032$ (See for details the
paper \cite{AK4}). }}
\end{figure}

 We have performed a numerical simulation of  the \index{Kramers' problem}Kramers' problem \cite{Cercignani}.  Kramers'  problem is a
limiting case of the plane Couette flow, where one of the plate is moved to infinity, while keeping a
fixed shear rate. We compare the analytical solution for the \index{Slip-velocity}slip-velocity at
the wall calculated for the  linearized BGK collision model \cite{Cercignani}  with the numerical
solution in the Fig. \ref{LBFig2}.

This shows that  one important feature of original Boltzmann equation, the Knudsen number dependent
slip at the wall is retained in the present model.

\begin{figure}[t]
\centering{
\includegraphics[scale=0.35]{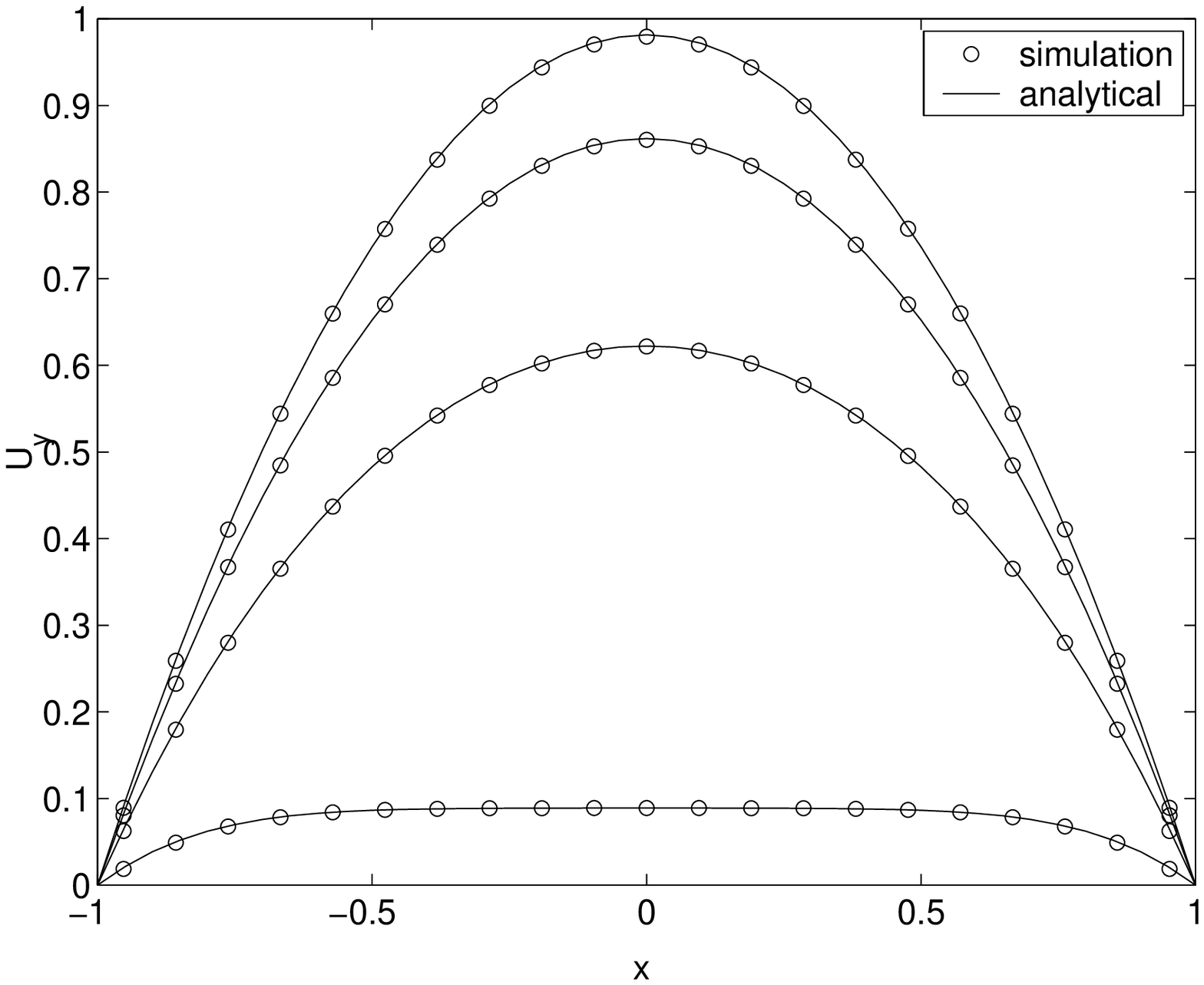}
\caption{\label{LBFig3} Development of the velocity profile in the Poiseuille flow. Reduced velocity
$U_y(x) = u_y/u_{y_{\rm max}}$ is shown versus the reduced coordinate across the channel $x$. Solid
line: Analytic solution. Different lines correspond to different instants of the reduced time $T=
(\nu t)/ ( 4 R^2)$, increasing from bottom to top. Here, $R$ is the half-width of the channel.
Symbol: simulation with the present ELBM algorithm. Parameters used are: viscosity $\nu= 5.0015
\times 10^{-5} (\beta = 0.9997)$, steady state maximal velocity
 $u_{y_{\rm max}} = 1.10217 \times 10^{-2} $. Reynolds number $Re =
1157$. (See for more details in the paper \cite{AKJSP}) }}
\end{figure}

In another numerical experiment, the method was  tested in the
 setup of the two-dimensional \index{Poiseuille flow}Poiseuille flow.
 The time evolution of the computed profile as compared to the
 analytical profile obtained from solving the Navier--Stokes equation is
demonstrated in  Fig.~\ref{LBFig3}.

\subsection{Outlook}
 We have constructed the minimal  kinetic models
for both athermal and thermal hydrodynamic simulations. The natural extension of the Gauss-Hermite
quadrature discretization towards the Boltzmann $H$ function  leads to physically sound kinetic
models consistent with the $H$ theorem.  A new approach to the discretization is taken, which ensures
non linear stability of the numerical method. The simulation of the Kramers' problem as a function of
the Knudsen number demonstrates the usefulness of the present approach for the low Knudsen number
flows.

\section{Other kinetic equations}

\subsection{The Enskog equation for hard spheres}

The Enskog equation for hard spheres is an extension of the Boltzmann equation to moderately dense
gases. The Enskog equation explicitly takes into account the nonlocality of collisions through a
two-fold modification of the Boltzmann collision integral: First, the one-particle distribution
functions are evaluated at the locations of the centers of spheres, separated by the nonzero distance
at the impact. This makes the collision integral nonlocal in space. Second, the equilibrium pair
distribution function at the contact of the spheres enhances the scattering probability.

Enskog's collision integral for hard spheres of radius $r_0$ is written in the following form
\cite{Chapman}:
\begin{eqnarray}\label{EnskEqua}
Q &=& \int_{R^3} \int_{B^-} \left[(\vv - \ww) \cdot \nn\right] \left[ \chi(\xx,\xx + r_0 \nn)f( \xx ,
\vv^{\prime}) f(\xx + 2r_0 \nn, \ww^{\prime}) \right. \nonumber
\\ && - \left. \chi(\xx, \xx- r_0 \nn)
 f(  \xx , \vv ) f( \xx - 2r_0 \nn , \ww ) \right] \, \D \ww   \, \D \nn,
\end{eqnarray}
where $\chi(\xx,\yy)$ is the equilibrium pair-correlation function for given temperature and density,
and integration in $\ww$ is carried over the whole space $R^3$, while integration in $\nn$ goes over
a hemisphere $B^-=\{\nn \in S^2 \mid (\ww-\vv,\nn)<0 \}$.

The proof of the $H$ theorem for the Enskog equation has posed certain difficulties, and has led to a
modification of the collision integral \cite{VanBeij}.

Methods of solution of the Enskog equation are immediate generalizations of those developed for the
Boltzmann equation, but there is one additional difficulty. The Enskog collision integral is nonlocal
in space. The Chapman-Enskog method, when applied to the Enskog equation, is supplemented with a
gradient expansion around the homogeneous equilibrium state.

\subsection{The Vlasov equation}

The Vlasov equation (or kinetic equation for a self-consistent force) is the nonlinear equation for
the one-body distribution function, which takes into account a long-range interaction between
particles:
\begin{eqnarray}
{\partial \over \partial t}f+\left(\vv,{\partial \over \partial \xx}f\right)+\left(\FF ,{\partial
\over
\partial \vv}f\right)=0,\nonumber
\end{eqnarray}
where  $\FF=\int \Phi(\mid \xx-\xx'\mid){\xx-\xx'\over \mid \xx-\xx'\mid}n(\xx') \,\D \xx'$ is the
self-consistent force. In this expression $ \Phi(\mid \xx-\xx'\mid){\xx-\xx'\over \mid \xx-\xx'\mid}$
is the microscopic force between the two particles, and $n(\xx')$ is the density of particles,
defined self-consistently, $n(\xx')=\int f(\xx',\vv) \,\D \vv.$

 The Vlasov equation is used for a description of collisionless plasmas in
which case it is completed by the set of Maxwell equation for the electromagnetic field \cite{LPi}.
It is also used for a description of the gravitating gas.

The Vlasov equation is an infinite-dimensional Hamiltonian system \cite{Marsd}. Many special and
approximate (wave-like) solutions to the Vlasov equation are known and they describe important
physical effects \cite{BraunHepp}. One of the most well known effects is the \index{Landau
damping}Landau damping \cite{LPi}: The energy of a volume element dissipates with the rate $$Q\approx
-\mid E\mid^2{\omega(k)\over k^2}\left.{\D f_0\over \D v}\right|_{v={\omega \over k}},$$ where $f_0$
is the Maxwell distribution function, $\mid E \mid$ is the amplitude of the applied monochromatic
electric field with the frequency $\omega(k)$ , $k$ is the wave vector. The Landau damping is
thermodynamically reversible effect, and it is not accompanied with an entropy increase.
Thermodynamically reversed to the Landau damping is the plasma echo effect.

\subsection{The Fokker-Planck equation}

The Fokker-Planck equation (FPE) is a familiar model in various problems of nonequilibrium
statistical physics \cite{VanKampen,Risk}. We consider the FPE of the form
\begin{equation}
\label{SFP} {\partial W(\xx,t) \over \partial t} ={\partial \over \partial \xx} \left\{D
\left[W{\partial \over
\partial \xx} U +{\partial \over \partial \xx} W\right]\right\}.
\end{equation}
Here $W(\xx,t)$ is the probability density over the configuration space $x$, at the time $t$, while
$U(\xx)$ and $D(\xx)$ are the potential and the positively semi-definite ($ (\yy,D\yy)\ge 0$)
diffusion matrix.

The FPE (\ref{SFP}) is particularly important in studies of polymer solutions \cite{Bird,Doi,HCO}.
Let us recall the two properties of the FPE (\ref{SFP}): (i). Conservation of the total probability:
$ \int W(\xx,t) \,\D  x\equiv 1.$ (ii). Dissipation: The equilibrium distribution,
$W_{eq}\propto\exp(-U)$, is the unique stationary solution to the FPE (\ref{SFP}). The entropy,
\begin{equation}
\label{entropy} S[W]=-\int W(\xx,t) \ln\left[\frac{W(\xx,t)}{W_{eq}(\xx)}\right] \,\D x,
\end{equation}
is a monotonically growing function due to the FPE (\ref{SFP}), and it arrives at the global maximum
in the equilibrium. These properties become most elicit when the FPE (\ref{SFP}) is rewritten as
follows:
\begin{equation}
\label{GENERIC}
\partial_{t}W(\xx,t)=\hat{M}_W\frac{\delta S[W]}{\delta W(\xx,t)},
\end{equation}
where $$\hat{M}_W=-{\partial \over \partial \xx} \left[W(\xx,t)D(\xx)  {\partial \over \partial \xx}
\right]$$ is a positive semi--definite symmetric operator with kernel $1$. The form (\ref{GENERIC})
is the dissipative part of a structure termed GENERIC (the dissipative vector field is a metric
transform of the entropy gradient) \cite{GENERIC,GENERIC1}.

The entropy does not depend on kinetic constants. It is the same for different details of kinetics,
and depends only on the equilibrium data. Let us call this property {\it ``universality"}. It is
known that for the Boltzmann equation there exists only one \index{Universal Lyapunov
functional}universal Lyapunov functional. It is the entropy (we do not distinguish functionals which
are related to reach other by monotonic transformation). But for the FPE there exists a big family of
universal Lyapunov functionals. Let $h(a)$ be a convex function of one variable $a\geq 0$,
$h''(a)>0$,

\begin{equation} \label{FPES}
S_h[W]=-\int W_{eq}(\xx) h\left[\frac{W(\xx,t)}{W_{eq}(\xx)}\right] \,\D x.
\end{equation}

The density of production of the generalized entropy $S_h$,  $\sigma_h$ is nonnegative:

\begin{equation} \label{sigmah}
\sigma_h(\xx)=W_{eq}(\xx)h''\left[\frac{W(\xx,t)}{W_{eq}(\xx)}\right]\left({\partial \over \partial
\xx}\frac{W(\xx,t)}{W_{eq}(\xx)},D{\partial \over \partial
\xx}\frac{W(\xx,t)}{W_{eq}(\xx)}\right)\geq 0.
\end{equation}

\index{Generalized entropy}The most important variants for the choice of $h$:

\noindent $h(a)=a\ln a$, $S_h$ is the \index{Boltzmann-Gibbs-Shannon
entropy}Boltzmann--Gibbs--Shannon entropy \index{Kullback entropy}(in the Kullback form
\cite{Kull,Pla}),

\noindent $h(a)=a\ln a-\epsilon \ln a$, $\epsilon>0$, $S_h^\epsilon$ is the maximal family of {\it
additive} entropies \cite{ENTR1,ENTR2,ENTR3} (these entropies are additive for composition of
independent subsystems).

\noindent $h(a)=\frac{1-a^q}{1-q}$, $S_h^q$ is the family of \index{Tsallis entropy}Tsallis entropies
\cite{Tsa,Abe}. These entropies are not additive, but become additive after nonlinear monotonous
transformation. This property can serve as a definition of the Tsallis entropies in the class of
generalized entropies (\ref{FPES}) \cite{ENTR3}.

\section{Equations of chemical kinetics and their reduction}

\subsection{Outline of the dissipative reaction kinetics} \label{KINETICS} We begin with an outline
of the reaction kinetics (for details see e.\ g.\ the book
 \cite{Yab}). Let us consider a closed system with $n$ chemical species ${\rm A}_1,\dots,{\rm
A}_n$, participating in a complex reaction. The complex reaction is represented by the following
\index{Stoichiometric equation}stoichiometric mechanism:
\begin{equation}
\label{stoi} \alpha_{s1}{\rm A}_1+\ldots+\alpha_{sn}{\rm A}_n\rightleftharpoons \beta_{s1}{\rm
A}_1+\ldots+\beta_{sn}{\rm A}_n,
\end{equation}
where the index $s=1,\dots,r$ enumerates the reaction steps, and where integers, $\alpha_{si}$ and
$\beta_{si}$,  are stoichiometric coefficients. For each reaction step $s$, we introduce
$n$--component vectors $\balpha_s$ and $\bbeta_s$ with components $\alpha_{si}$ and $\beta_{si}$.
Notation $\mbox{\boldmath$\gamma$}_s$  stands for the vector with integer components
$\gamma_{si}=\beta_{si}-\alpha_{si}$ \index{Stoichiometric vector}(the stoichiometric vector).

For every $A_i$ an {\it extensive variable} $N_i$, ``the number of particles of i-th specie", is
introduced. The concentration of $A_i$ is $c_i=N_i/V$, where $V$ is the volume.

Given the stoichiometric mechanism (\ref{stoi}), the reaction kinetic equations read:
\begin{equation}\label{reaction}
\dot{\NN}= V\JJ(\cc),\ \JJ(\cc)=\sum_{s=1}^{r}\bgamma_sW_s(\cc),
\end{equation}
where dot denotes the time derivative, and $W_s$  is the reaction rate function of the step $s$. In
particular, \index{Mass action law} {\it the mass action law} suggests the polynomial form of the
reaction rates:
\begin{equation}
\label{MAL} W_s(\cc)=W_s^+(\cc) -  W_s^-(\cc) = k^+_s(T) \prod_{i=1}^{n}c_i^{\alpha_{si}} - k^-_s(T)
\prod_{i=1}^{n} c_i^{\beta_{si}},
\end{equation}
where $k^+_s(T)$ and $k^-_s(T)$ are the constants of the direct and of the inverse reactions rates of
the $s$th reaction step, $T$ is the temperature. The (generalized)\index{Arrhenius equation}Arrhenius
equation gives the most popular form of dependence $k^+_s(T)$:

\begin{equation}\label{Arr}
k^{\pm}_s(T)=a^{\pm}_s T^{b^{\pm}_s} \exp(S^{\pm}_s/k_{\rm B}) \exp(-H^{\pm}_s/k_{\rm B}T),
\end{equation}
\noindent where $a^{\pm}_s, \: b^{\pm}_s$ are constants, $H^{\pm}_s$ are activation ethalpies,
$S^{\pm}_s$ are activation entropies.

The rate constants are not independent. The \index{Detailed balance}{\it principle of detailed
balance} gives the following connection between these constants: There exists such a positive vector
$\cc^{\rm eq}(T)$ that

\begin{equation}\label{dbchem}
W_s^+(\cc^{\rm eq})=W_s^-(\cc^{\rm eq}) \: \mbox{for all}  \: s=1,\dots,r .
\end{equation}

The necessary and sufficient conditions for existence of such $\cc^{\rm eq}$ can be formulate as the
system of polynomial equalities for $\{k^{\pm}_s\}$, if the the stoichiometric vectors
$\{\bgamma_{s}\}$ are linearly dependent (see, for example, \cite{Yab}).

The reaction kinetic equations (\ref{reaction}) do not give us a closed system of equations, because
dynamics of the volume $V$ is not defined yet. Four classical conditions for closure of this system
are well studied: $U,\: V = {\rm const}$ (isolated system, $U$ is the internal energy); $H,\: P =
{\rm const}$ (thermal isolated isobaric system, $P$ is the pressure, $H=U+PV$ is the enthalpy), $V,
\: T = {\rm const}$ (isochoric isothermal conditions); $P, \: T = {\rm const}$ (isobaric isothermal
conditions). For $V, \: T = {\rm const}$ we do not need additional equations and data. It is possible
just to divide equation (\ref{reaction}) by the constant volume and write

\begin{equation}\label{creaction}
\dot{\cc}= \sum_{s=1}^{r}\bgamma_sW_s(\cc).
\end{equation}

For non-isothermal and non-isochoric conditions we do need addition formulae to derive $T$ and $V$.
For all the four classical conditions the \index{Thermodynamic Lyapunov functions}thermodynamic
Lyapunov functions $G$ for kinetic equations are known:
\begin{eqnarray}\label{tdlya}
&& U,\: V = {\rm const}, \: G_{U,V}=-S/k_{\rm B}; \nonumber \\ && V, \: T = {\rm const}, \:
G_{V,T}=F/k_{\rm B}T=U/k_{\rm B}T-S/k_{\rm B}; \nonumber \\ && H,\: P = {\rm const}, \:
G_{H,P}=-S/k_{\rm B}; \nonumber \\ && P, \: T = {\rm const}, \: G_{P,T}=G/k_{\rm B}T=H/k_{\rm
B}T-S/k_{\rm B},
\end{eqnarray}
\noindent where $F=U-TS$ is the \index{Free energy}free energy (Helmholtz free energy), $G=H-TS$ is
the \index{Free entalphy}free entalphy (Gibbs free energy). All the thermodynamic Lyapunov functions
are normalized to the dimensionless scale (if one measures the number of particles in moles, then it
is necessary to change $k_{\rm B}$ to $R$). All these functions decrease in time. For classical
conditions the corresponding thermodynamic Lyapunov functions can be written in the form:
$G_{\bullet}({\rm const}, \NN)$. The derivatives $\partial G_{\bullet}({\rm const}, \NN) / \partial
N_i$ are the same functions of $\cc$ and $T$ for all classical conditions:
\begin{equation} \label{mu}
\mu_i(\cc,T)=\frac{\partial G_{\bullet}({\rm const}, \NN)}{\partial N_i}=\frac{\mu_{{\rm chem}
i}(\cc,T)}{k_{\rm B}T},
\end{equation}
where $\mu_{{\rm chem} i}(\cc,T)$ is the chemical potential of $A_i$.

Usual $G_{\bullet}({\rm const}, \NN)$ are strictly convex functions of $\NN$, and the matrix
$\partial \mu_i / \partial c_j$ is positively definite. The dissipation inequality (\ref{Htheorem})
holds
\begin{equation} \label{Htheorem}
\frac{ \,\D G_{\bullet}}{ \,\D t}=V(\bmu,\JJ) \leq 0.
\end{equation}
This inequality is the restriction on  possible kinetic law and on possible values of kinetic
constants.


One of the most important generalization of the mass action law (\ref{MAL}) is  the
\index{Marcelin-De Donder kinetics}Marcelin-De Donder kinetic function. This generalization
\cite{Fein,ByGoYa} is based on ideas of the thermodynamic theory of affinity \cite{DeDonder36}. We
use the kinetic function suggested in its final form in \cite{ByGoYa}. Within this approach, the
functions $W_s$ are constructed as follows: For a given $\bmu(\cc,T)$ (\ref{mu}), and for a given
stoichiometric mechanism (\ref{stoi}), we define the gain ($+$) and the loss ($-$) rates of the $s$th
step,
\begin{equation}
\label{MDD} W_s^{+}=\varphi_s^+ \exp( \bmu,\mbox{\boldmath$\alpha$}_s),\quad W_s^{-}=\varphi_s^-\exp(
\bmu,\mbox{\boldmath$\beta$}_s),
\end{equation}
where $\varphi_s^{\pm}>0$ are kinetic factors, $(\:,\:)$ is the standard inner product, the sum of
coordinates  products.

The Marcelin-De Donder kinetic function reads: $W_s=W_s^+-W_s^-$, and the right hand side of the
kinetic equation (\ref{reaction}) becomes,
\begin{equation}
\label{KINETIC MDD} \JJ=\sum_{s=1}^{r}\bgamma_s \{\varphi_s^+ \exp(\bmu,\balpha_s)-
\varphi_s^-\exp(\bmu,\bbeta_s)\}.
\end{equation}
For the Marcelin-De Donder reaction rate (\ref{MDD}), the dissipation inequality (\ref{Htheorem}) is
particularly  elegant:
\begin{equation}
\label{HMDD} \dot{G}=\sum_{s=1}^{r} [(\bmu,\bbeta_s) - (\bmu,\balpha_s)] \left\{\varphi_s^+
e^{(\mbox{\boldmath {\scriptsize $\mu$}},\mbox{\boldmath {\scriptsize $\alpha$}}_s)}-
\varphi_s^-e^{(\mbox{\boldmath {\scriptsize $\mu$}},\mbox{\boldmath {\scriptsize
$\beta$}}_s)}\right\}\le 0.
\end{equation}
The kinetic factors $\varphi_s^{\pm}$ should satisfy certain conditions in order to make valid the
dissipation inequality (\ref{HMDD}). A well known sufficient condition is the detailed balance:
\begin{equation}
\label{DB} \varphi_s^+=\varphi_s^-,
\end{equation}
other sufficient conditions are discussed in detail elsewhere \cite{G1,Yab,DKN97}.

For ideal systems, function $G_{\bullet}$ is constructed from the thermodynamic data of individual
species. It is convenient to start from the isochoric isothermal conditions. The Helmholtz free
energy for the ideal system is
\begin{equation}
\label{Freen} F=k_{\rm B}T \sum_i N_i[\ln c_i - 1 + \mu_{0i}]+{\rm const}_{T,V},
\end{equation}
where the internal energy is assumed to be a linear function of $N$ in a given interval of $\cc$,
$T$: $$U=\sum_i N_iu_i(T)=\sum_i N_i (u_{0i} + C_{Vi}T),$$ where $u_i(T)$ is the internal energy of
$A_i$ per particle. It is well known that $S=-(\partial F/ \partial T)_{V,N={\rm const}}$,
$U=F+TS=F-T(\partial F/
\partial T)_{V,N={\rm const}}$, hence, $u_i(T)=-k_{\rm B}T^2d\mu_{0i}/dT$ and
\begin{equation}
\label{mu0} \mu_{0i}= \delta_i + u_{0i}/k_{\rm B}T - (C_{Vi}/k_{\rm B})\ln T,
\end{equation}
where $\delta_i={\rm const}$, $C_{Vi}$ is the $A_i$ heat capacity at constant volume (per particle).

In concordance with the form of ideal free energy (\ref{Freen}) the expression for $\bmu$ is:
\begin{equation}
\label{muid} \mu_{i}=  \ln c_i +  \delta_i + u_{0i}/k_{\rm B}T - (C_{Vi}/k_{\rm B})\ln T.
\end{equation}

For the function $\bmu$ of the form (\ref{muid}), the Marcelin-De Donder equation casts into the more
familiar mass action law form (\ref{MAL}). Taking into account the principle of detailed balance
(\ref{DB}) we get the ideal rate functions:
\begin{eqnarray}
\label{MALMD} W_s(\cc)&=&W_s^+(\cc)-W_s^-(\cc), \nonumber \\ W_s^+(\cc)&=&\varphi_s(\cc,T) T^{-\sum_i
\alpha_{si}C_{Vi}/k_{\rm B}} e^{\sum_i \alpha_{si}(\delta_i + u_{0i}/k_{\rm
B}T)}\prod_{i=1}^{n}c_i^{\alpha_{si}}, \nonumber \\ W_s^-(\cc) &=&\varphi_s(\cc,T) T^{-\sum_i
\beta_{si}C_{Vi}/k_{\rm B}} e^{\sum_i \beta_{si}(\delta_i + u_{0i}/k_{\rm B}T)}\prod_{i=1}^{n}
c_i^{\beta_{si}}.
\end{eqnarray}
where $\varphi_s(\cc,T)$ is an arbitrary positive function (from thermodynamic point of view).

Let us discuss further the vector field $\JJ(\cc)$ in the concentration space (\ref{creaction}).
Conservation laws (balances) impose linear constraints on admissible vectors $ \D \cc/ \D t$:
\begin{equation}
\label{conser} (\bb_i, \cc)=B_i={\rm const}, \: \left(\bb_i, {\D \cc \over \D t}\right)=0, \:
i=1,\dots,l,
\end{equation}
where $\bb_i$ are fixed and linearly independent vectors. Let us denote as $\BB$ the set of vectors
which satisfy the conservation laws (\ref{conser}) with given $B_i$:
\[
\BB=\left\{\cc|(\bb_1,\cc)=B_1,\dots, (\bb_l,\cc)=B_l\right\}.
\]
The natural phase space $\XX$ of the system (\ref{creaction}) is the intersection of the cone of
$n$-dimensional vectors with nonnegative components, with the set $\BB$, and ${\rm dim}\XX=d=n-l$. In
the sequel, we term a vector $\cc\in\XX$ the state of the system. In addition, we assume that each of
the conservation laws is supported by each elementary reaction step, that is
\begin{equation}
\label{sep} (\bgamma_s,\bb_i)=0,
\end{equation}
for each pair of vectors $\bgamma_s$ and $\bb_i$.

Reaction kinetic equations describe variations of the states in time.  The phase space $\XX$ is
positive-invariant of the system (\ref{creaction}): If $\cc(0)\in\XX$, then $\cc(t)\in\XX$ for all
the times $t>0$.

In the sequel, we assume that the kinetic equation (\ref{creaction}) describes  evolution towards the
unique equilibrium state, $\cc^{\rm eq}$, in the interior of the phase space $\XX$. Furthermore, we
assume that there exists a strictly convex function $G(\cc)$ which decreases monotonically in time
due to (\ref{creaction}):

Here $\bnabla G$ is the vector of partial derivatives $\partial G/\partial c_i$, and the convexity
assumes that the $n\times n$ matrices
\begin{equation}
\label{MATRIX} \HH_{{\mbox{\boldmath {\scriptsize $\cc$}}}}=\|\partial^2G(\cc)/\partial c_i\partial
c_j\|,
\end{equation}
are positive definite for all $\cc\in\XX$. In addition, we assume that the matrices (\ref{MATRIX})
are invertible if $\cc$ is taken in the interior of the phase space.

The function $G$ is the Lyapunov function of the system (\ref{reaction}), and $\cc^{\rm eq}$ is the
point of global minimum of the function $G$ in the phase space $\XX$. Otherwise stated, the manifold
of equilibrium states $\cc^{\rm eq}(B_1,\dots,B_l)$ is the solution to the variational problem,
\begin{equation}
\label{EQUILIBRIUM} G\to{\rm min}\ {\rm for\ }(\bb_i,\cc)=B_i,\ i=1,\dots,l.
\end{equation}
For each fixed value of the conserved quantities $B_i$, the solution is unique. In many cases,
however, it is convenient to consider the whole equilibrium manifold, keeping the conserved
quantities as parameters.

For example, for perfect systems in a constant volume under a constant temperature, the Lyapunov
function $G$ reads:
\begin{equation}
\label{gfun} G=\sum_{i=1}^{n}c_i[\ln(c_i/c^{\rm eq}_i)-1].
\end{equation}

It is important to stress that $\cc^{\rm eq}$ in (\ref{gfun}) is an {\it arbitrary} equilibrium of
the system, under arbitrary values of the balances. In order to compute $G(\cc)$, it is unnecessary
to calculate the specific equilibrium $\cc^{\rm eq}$ which corresponds to the initial state $\cc$.
Let us compare the Lyapunov function $G$ (\ref{gfun}) with the classical formula for the free energy
(\ref{Freen}). This comparison gives a possible choice for $\cc^{\rm eq}$:
\begin{equation}
\ln c^{\rm eq}_i = -\delta_i - u_{0i}/k_{\rm B}T + (C_{Vi}/k_{\rm B})\ln T.
\end{equation}

\subsection{The problem of reduced description in chemical kinetics}

\label{reduction_review} What does it mean, ``to reduce the description of a chemical system''? This
means the following:
\begin{enumerate}
\item To shorten  the list of species.
This, in turn, can be achieved in two ways:

(i) To eliminate inessential components from the list;

(ii) To lump some of the species into integrated components.

\item To shorten the list of reactions. This also can be done in several ways:

(i) To eliminate inessential reactions, those which do not significantly influence the reaction
process;

(ii) To assume that some of the reactions ``have been already completed'', and that the equilibrium
has been reached along their paths (this leads to dimensional reduction because the rate constants of
the ``completed'' reactions are not used thereafter, what one needs are equilibrium constants only).

\item To decompose the motions into fast and slow, into independent (almost-independent)
and slaved etc. As the result of such a decomposition, the system admits a study ``in parts''. After
that, results of this study are combined into a joint picture. There are several approaches which
fall into this category. The famous method of the {\it quasi-steady state} (QSS), pioneered by
Bodenstein and Semenov,  follows the Chapman-Enskog method. The {\it partial equilibrium
approximations} are predecessors of the Grad method and quasiequilibrium approximations in physical
kinetics. These two family of methods have different physical backgrounds and mathematical forms.

\end{enumerate}

\subsection{Partial equilibrium approximations}\label{partial_eq}

{\it Quasiequilibrium with respect to reactions} is constructed as follows: From the list of
reactions (\ref{stoi}), one selects those which are assumed to equilibrate first. Let they be indexed
with the numbers $s_1,\dots,s_k$. The quasiequilibrium manifold is defined by the system of
equations,
\begin{equation}
\label{st1} W^+_{s_i}=W^-_{s_i},\ i=1,\dots,k.
\end{equation}
This system of equations looks particularly elegant when written in terms of conjugated (dual)
variables,  $\bmu=\bnabla G$:
\begin{equation}
\label{st2} ( \bgamma_{s_i},\bmu)=0,\ i=1,\dots,k.
\end{equation}
In terms of conjugated variables, the quasiequilibrium manifold forms a linear subspace. This
subspace, $L^{\perp}$, is the orthogonal completement to the linear envelope of vectors, $L={\rm
lin}\{\bgamma_{s_1},\dots,\bgamma_{s_k}\}$.

{\it Quasiequilibrium with respect to species} is constructed practically  in the same way but
without selecting the subset of reactions. For a given set of species, $A_{i_1}, \dots, A_{i_k}$, one
assumes that their concentrations evolve fast to the equilibrium, and remain there. Formally, this
means that in the $k$-dimensional subspace of the space of concentrations with the coordinates
$c_{i_1},\dots,c_{i_k}$, one constructs the subspace $L$ which is defined by the balance equations,
$( \bb_i,\cc)=0$. In terms of the conjugated variables, the quasiequilibrium manifold, $L^{\perp}$,
is defined by equations,
\begin{equation}
\label{qe1} \bmu\in L^{\perp},\ (\bmu=(\mu_1,\dots,\mu_n)).
\end{equation}
The same quasiequilibrium manifold can be also defined with the help of fictitious reactions: Let
$\bg_1,\dots,\bg_q$ be a basis in $L$. Then (\ref{qe1}) may be rewritten as follows:
\begin{equation}
\label{qe2} ( \bg_i,\bmu)=0,\ i=1,\dots,q.
\end{equation}

{\it Illustration:} Quasiequilibrium with respect to reactions in hydrogen oxidation: Let us assume
equilibrium with respect to dissociation reactions, ${\rm H}_2\rightleftharpoons 2{\rm H}$, and,
${\rm O}_2\rightleftharpoons 2{\rm O}$, in some subdomain of reaction conditions. This gives:
\[k_1^+c_{{\rm H}_2}=k_1^-c^2_{\rm H},\ k_2^+c_{{\rm O}_2}=k_2^-c_{\rm O}^2.\]
Quasiequilibrium with respect to species: For the same reaction, let us assume equilibrium over ${\rm
H}$, ${\rm O}$, ${\rm OH}$, and ${\rm H}_2{\rm O}_2$, in a subdomain of reaction conditions. Subspace
$L$ is defined by balance constraints:
\[ c_{\rm H}+c_{\rm OH}+2c_{{\rm H}_2{\rm O}_2}=0,\ c_{\rm O}+c_{\rm OH}
+2c_{{\rm H}_2{\rm O}_2}=0.\] Subspace $L$ is two-dimensional. Its basis, $\{\bg_1,\bg_2\}$ in the
coordinates $c_{\rm H}$, $c_{\rm O}$, $c_{\rm OH}$, and $c_{{\rm H}_2{\rm O}_2}$ reads:
\[
\bg_1=(1,1,-1,0),\quad \bg_2=(2,2,0,-1).
\]
Corresponding (\ref{qe2}) is:
\[ \mu_{\rm H}+\mu_{\rm O}=\mu_{\rm OH},\ 2\mu_{\rm H}+2\mu_{\rm O}=
\mu_{{\rm H}_2{\rm O}_2}.\]

\index{Quasiequilibrium!manifold}{\it General construction of the quasiequilibrium manifold}: In the
space of concentration, one defines a subspace $L$ which satisfies the balance constraints:
\[ ( \bb_i,L)\equiv0.\]
The orthogonal complement of $L$ in the space with coordinates $\bmu=\bnabla G$ defines then the
quasiequilibrium manifold $\bOmega_{L}$. For the actual computations, one requires the inversion from
$\bmu$ to $\cc$. Duality structure $\bmu\leftrightarrow\cc$ is well studied by many authors
\cite{Orlov84,DKN97}.

\index{Quasiequilibrium!projector}{\it Quasiequilibrium projector.} It is not sufficient to just
derive the manifold, it is also required to define a {\it projector} which would transform the vector
field defined on the space of concentrations into a vector field on the manifold. Quasiequilibrium
manifold consists of points which minimize $G$ on the affine spaces of the form $\cc+L$. These affine
planes are hypothetic planes of fast motions ($G$ is decreasing in the course of the fast motions).
Therefore, the quasiequilibrium projector maps the whole space of concentrations on $\bOmega_L$
parallel to $L$. The vector field is also projected onto the tangent space of $\bOmega_L$ parallel to
$L$.

Thus, the quasiequilibrium approximation assumes the decomposition of motions into the fast -
parallel to $L$, and the slow - along the quasiequilibrium manifold. In order to construct the
quasiequilibrium approximation, knowledge of reaction rate constants of ``fast'' reactions is not
required (stoichiometric vectors of all these fast reaction are in $L$, $\bgamma_{{\rm fast}}\in L$,
thus, the knowledge of $L$ suffices), one only needs some confidence in that they all are
sufficiently fast \cite{Volpert85}. The quasiequilibrium manifold itself is constructed based on the
knowledge of $L$ and of $G$. Dynamics on the quasiequilibrium manifold is defined as the
quasiequilibrium projection of the ``slow component'' of kinetic equations (\ref{reaction}).

\subsection{Model equations}

The assumption behind  the quasiequilibrium is the hypothesis of the decomposition of motions into
fast and slow. The quasiequilibrium approximation itself describes slow motions. However, sometimes
it becomes necessary to restore to the whole system, and to take into account the fast motions as
well. With this, it is desirable to keep intact one of the important advantages of the
quasiequilibrium approximation -
 its independence of the rate constants of fast reactions.
For this purpose, the detailed fast kinetics is replaced by a model equation ({\it single relaxation
time approximation}).

\index{Quasiequilibrium!models}{\it Quasiequilibrium models} (QEM) are constructed as follows: For
each concentration vector $\cc$, consider the affine manifold, $\cc+L$. Its intersection with the
quasiequilibrium manifold $\bOmega_L$ consists of one point. This point delivers the  minimum to $G$
on $\cc+L$. Let us denote this point as $\cc^*_L(\cc)$. The equation of the quasiequilibrium model
reads:
\begin{equation}
\label{QEmodel} \dot{\cc}=-\frac{1}{\tau}[\cc-\cc^*_L(\cc)]+\sum_{{\rm
slow}}\bgamma_{s}W_s(\cc^*_L(\cc)),
\end{equation}
where $\tau>0$ is the relaxation time of the fast subsystem. Rates of slow reactions are computed at
the points $\cc^*_L(\cc)$ (the second term in the right hand side of (\ref{QEmodel}), whereas the
rapid motion is taken into account by a simple relaxational term (the first term in the right hand
side of (\ref{QEmodel}). The most famous model kinetic equation is the BGK equation in the theory of
the Boltzmann equation \cite{BGK}. The general theory of the quasiequilibrium models, including
proofs of their thermodynamic consistency, was constructed in the paper \cite{GKMod}.

\index{Gradient models}{\it Single relaxation time gradient models} (SRTGM) were introduced in the
context of the lattice Boltzmann method for hydrodynamics \cite{AKJSP,AKMod}. These models are aimed
at improving the obvious drawback of quasiequilibrium models (\ref{QEmodel}): In order to construct
the QEM, one needs to compute the function,
\begin{equation}
\label{QEA}
 \cc^*_L(\cc)=\arg\min_{\mbox{\boldmath {\scriptsize $x$}}\in
 \mbox{\boldmath {\scriptsize $c$}}+L,\ \mbox{\boldmath {\scriptsize $x$}}>0}G(\xxx).
\end{equation}
This is a convex programming problem. It does not always have a closed-form solution.

Let $\bg_1,\dots,\bg_k$ is some orthonormal basis of $L$. We denote as $\DD(\cc)$ the $k\times k$
matrix with the elements $( \bg_i,\HH_{\mbox{\boldmath {\scriptsize $\cc$}}} \bg_j)$, where
$\HH_{\mbox{\boldmath {\scriptsize $\cc$}}}$ is the matrix of second derivatives of $G$
(\ref{MATRIX}). Let $\CC(\cc)$ be the inverse of $\DD(\cc)$. The single relaxation time gradient
model has the form:
\begin{equation}
\dot{\cc}=-\frac{1}{\tau}\sum_{i,j=1}^k\bg_i\CC(\cc)_{ij}( \bg_j,\bnabla G) +\sum_{{\rm
slow}}\bgamma_{s}W_s(\cc).\label{SRTGM}
\end{equation}
The first term drives the system to the minimum of $G$ on $\cc+L$, it does not require solving the
problem (\ref{QEA}), and its spectrum in the quasiequilibrium is the same as in the quasiequilibrium
model (\ref{QEmodel}). Note that the slow component is evaluated in the ``current'' state $\cc$.

The first term of equation (\ref{SRTGM}) has a simple form
\begin{equation} \label{gradmodchem}
\dot{\cc}=-\frac{1}{\tau} {\rm grad} G+\sum_{{\rm slow}}\bgamma_{s}W_s(\cc),
\end{equation}
if one calculates the gradient ${\rm grad} G \in L$ on the plane of fast motions $\cc+L$ with the
entropic scalar product \footnote{Let us remind that ${\rm grad} G$ is the \index{Riesz
representation}Riesz representation of the differential of $G$ in the phase space $\XX$: $G(\cc +
\Delta\cc)=G(\cc)+ \langle {\rm grad} G(\cc),$ $\Delta\cc \rangle + o(\Delta\cc)$. It belongs to the
tangent space of $\XX$ and depends on the scalar product. From thermodynamic point of view there is
only one distinguished scalar product in concentration space, the entropic scalar product. Usual
definition of ${\rm grad} G$ as the vector of partial derivatives corresponds to the standard scalar
product $(\bullet,\bullet)$ and to the choice: $\XX$ is the whole concentration space. In equation
(\ref{gradmodchem}) $\XX=\cc+L$ and we use the entropic scalar product.} $\langle \xx , \yy
\rangle=(\xx,\HH_{\mbox{\boldmath {\scriptsize $\cc$}}}\yy)$.

The models (\ref{QEmodel}) and (\ref{SRTGM})
 lift the quasiequilibrium approximation to a kinetic equation
by approximating the fast dynamics with a single ``reaction rate constant'' - relaxation time $\tau$.

\subsection{Quasi-steady state approximation}\label{QSS} The quasi-steady state approximation (QSS)
is a tool used in a major number of works. Let us split the list of species in two groups: The basic
and the intermediate (radicals etc). Concentration vectors are denoted accordingly, $\cc^{\rm s}$
(slow, basic species), and $\cc^{\rm f}$ (fast, intermediate species). The concentration vector $\cc$
is the direct sum, $\cc=\cc^{\rm s}\oplus\cc^{\rm f}$. The fast subsystem is  (\ref{reaction}) for
the component $\cc^{\rm f}$ at fixed values of $\cc^{{\rm s}}$. If it happens that in this way
defined fast subsystem relaxes to a stationary state, $\cc^{\rm f}\to\cc^{\rm f}_{\rm qss}(\cc^{\rm
s})$, then the assumption that $\cc^{\rm f}=\cc^{\rm f}_{\rm qss}(\cc)$ is precisely  the QSS
assumption. The slow subsystem is the part of the system (\ref{reaction}) for $\cc^{\rm s}$, in the
right hand side of which the component $\cc^{\rm f}$ is replaced with $\cc^{\rm f}_{\rm qss}(\cc)$.
Thus, $\JJ=\JJ_{\rm s}\oplus\JJ_{\rm f}$, where
\begin{eqnarray}
\dot{\cc}^{\rm f}&=&\JJ_{\rm f}(\cc^{\rm s}\oplus\cc^{\rm f}), \ \cc^{\rm s}={\rm const}; \quad
\cc^{\rm f}\to\cc^{\rm f}_{\rm qss}(\cc^{\rm s});\label{fast}\\ \dot{\cc}^{\rm s}&=&\JJ_{\rm
s}(\cc^{\rm s}\oplus\cc_{\rm qss}^{\rm f}(\cc^{\rm s})). \label{slow}
\end{eqnarray}
Bifurcations in the system (\ref{fast}) under variation of $\cc^{\rm s}$ as a parameter are
confronted to kinetic critical phenomena. Studies of more complicated dynamic phenomena in the fast
subsystem (\ref{fast}) require various techniques of averaging, stability analysis of the averaged
quantities etc.

Various versions of the QSS method are well possible, and are actually used widely, for example, the
hierarchical QSS method. There, one defines not a single fast subsystem but a hierarchy of them,
$\cc^{{\rm f}_1},\dots,\cc^{{\rm f}_k}$. Each subsystem  $\cc^{{\rm f}_i}$ is regarded as a slow
system for all the foregoing subsystems, and it is regarded as a fast subsystem for the following
members of the hierarchy. Instead of one system of equations (\ref{fast}), a hierarchy of systems of
lower-dimensional equations is considered, each of these subsystem is easier to study analytically.

Theory of singularly perturbed systems of ordinary differential equations is used to provide a
mathematical background and refinements of the QSS approximation. In spite of a broad literature on
this subject, it remains, in general, unclear, what is the smallness parameter that separates the
intermediate (fast) species from the basic (slow). Reaction rate constants cannot be such a parameter
(unlike in the case of the quasiequilibrium). Indeed, intermediate species participate in the {\it
same} reactions, as the basic species (for example, ${\rm H}_2\rightleftharpoons 2{\rm H}$, ${\rm
H}+{\rm O}_2\rightleftharpoons {\rm OH}+{\rm O}$). It is therefore incorrect to state that $\cc^{\rm
f}$ evolve faster than $\cc^{\rm s}$. In the sense of reaction rate constants, $\cc^{\rm f}$ is not
faster.

For catalytic reactions, it is not difficult to figure out what is the smallness parameter that
separates the intermediate species from the basic, and which allows to upgrade the QSS assumption to
a singular perturbation theory rigorously \cite{Yab}. This smallness parameter is the ratio of
balances: Intermediate species include a catalyst, and their total amount is simply significantly
smaler than the amount of all the $\cc_i$'s. After renormalizing to the variables of one order of
magnitude, the small parameter appears explicitly. The simplest example provides the catalytic
reaction $A+Z\rightleftharpoons AZ \rightleftharpoons P+Z$ (here $Z$ is a catalyst, $A$ and $P$ are
an initial substrate and a product). The kinetic equations are (in obvious notations):
\begin{eqnarray}\label{MihMen}
\dot{c}_A&=&-k^+_1 c_A c_Z + k^-_1 c_{AZ}, \nonumber \\ \dot{c}_Z &=& -k^+_1 c_A c_Z + k^-_1 c_{AZ} +
k^+_2 c_{AZ} - k^-_2 c_{Z}c_P, \nonumber \\ \dot{c}_{AZ}&=&k^+_1 c_A c_Z - k^-_1 c_{AZ} - k^+_2
c_{AZ} + k^-_2 c_{Z}c_P, \nonumber \\ \dot{c}_P&=&k^+_2 c_{AZ} - k^-_2 c_{Z}c_P.
\end{eqnarray}
The constants and the reactions rates are the same for concentrations $c_A, c_P$, and for $c_Z,
c_{AZ}$, and they cannot be a reason for the relative slowness of $c_A, c_P$ in comparison with $c_Z,
c_{AZ}$.  However, there may be another source of slowness. There are two balances for this kinetics:
$c_A+c_P+c_{AZ}=B_A,$ $c_Z+c_{AZ}=B_Z$. Let us switch to the dimensionless variables:
$$\varsigma_A=c_A/B_A, \: \varsigma_P=c_P/B_A, \: \varsigma_Z=c_Z/B_Z, \:
\varsigma_{AZ}=c_{AZ}/B_Z.$$ The kinetic system (\ref{MihMen}) is then rewritten as
\begin{eqnarray}\label{MihMenDL}
\dot{\varsigma_A}&=&B_Z\left[-k^+_1 \varsigma_A \varsigma_Z + \frac{k^-_1}{B_A}\varsigma_{AZ}\right],
\nonumber \\ \dot{\varsigma_Z}&=&B_A\left[ -k^+_1 \varsigma_A \varsigma_Z +
\frac{k^-_1}{B_A}\varsigma_{AZ} + \frac{k^+_2}{B_A}\varsigma_{AZ} - k^-_2 \varsigma_Z \varsigma_P
\right], \nonumber \\ && \varsigma_A+\varsigma_P + \frac{B_Z}{B_A}\varsigma_{AZ}=1, \:
\varsigma_Z+\varsigma_{AZ}=1; \: \varsigma_{\bullet}\geq 0.
\end{eqnarray}
For $B_Z \ll B_A$ (the total amount of the catalyst is much smaller than the total amount of the
substrate) the slowness of $\varsigma_A, \: \varsigma_P$ is evident from these equations
(\ref{MihMenDL}).

 For usual radicals, the origin of the smallness parameter is quite similar. There are much
less radicals than the basic species (otherwise, the QSS assumption is inapplicable). In the case of
radicals, however, the smallness parameter cannot be extracted directly from balances $B_i$
(\ref{conser}). Instead, one can come up with a thermodynamic estimate: Function $G$ decreases in the
course of reactions, whereupon we obtain the limiting estimate of concentrations of any specie:
\begin{equation}
\label{TDlim} c_i\le \max_{G(\mbox{\boldmath {\scriptsize $c$}})\le G(\mbox{\boldmath {\scriptsize
$c$}}(0))} c_i,
\end{equation}
where $\cc(0)$ is the initial composition. If the concentration $c_{\rm R}$ of the radical R is small
both initially and in the equilibrium, then it should remain small also along the path to the
equilibrium. For example, in the case of ideal $G$ (\ref{gfun}) under relevant conditions, for any
$t>0$, the following inequality is valid:
\begin{equation}
\label{INEQ_R} c_{\rm R}[\ln(c_{\rm R}(t)/c_{\rm R}^{\rm eq})-1]\le G(\cc(0)).
\end{equation}
Inequality (\ref{INEQ_R}) provides the simplest (but rather crude) thermodynamic estimate of $c_{\rm
R}(t)$ in terms of $G(\cc(0))$ and $c_{\rm R}^{\rm eq}$ {\it uniformly for $t>0$}. Complete theory of
thermodynamic estimates of dynamics has been developed in the book \cite{G1}.

One can also do computations without a priori estimations, if one accepts the QSS assumption until
the values $\cc^{\rm f}$ stay sufficiently small. It is the simplest way to operate with QSS: Just
use it {\it until} $\cc^{\rm f}$ are small!

Let us assume that an a priori estimate has been found, $c_i(t)\le c_{i\ {\rm max}}$, for each $c_i$.
These estimate may depend on the initial conditions, thermodynamic data etc. With these estimates, we
are able to renormalize the variables in the kinetic equations (\ref{reaction}) in such a way that
the renormalized variables take their values from the unit segment $[0,1]$: $\tilde{c}_i=c_i/c_{i\
{\rm max}}$. Then the system (\ref{reaction}) can be written as follows:
\begin{equation}
\label{reduced} \frac{ \,\D \tilde{c}_i}{ \,\D t}=\frac{1}{c_{i\ {\rm max}}}J_i(\cc).
\end{equation}
The system of dimensionless parameters, $\epsilon_i=c_{i\ {\rm max}}/\max_i c_{i\ {\rm max}}$ defines
a hierarchy of relaxation times, and with its help one can establish various realizations of the QSS
approximation. The simplest version is the standard QSS assumption: Parameters $\epsilon_i$ are
separated in two groups, the smaller ones, and of the order $1$. Accordingly, the concentration
vector is split into $\cc^{\rm s}\oplus\cc^{\rm f}$. Various hierarchical QSS are possible, with
this, the problem becomes more tractable analytically.

There exists a variety of ways to introduce the smallness parameter into kinetic equations, and one
can find applications to each of the realizations. However, the two particular realizations remain
basic for chemical kinetics: (i) Fast reactions (under a given thermodynamic data); (ii) Small
concentrations. In the first case, one is led to the quasiequilibrium approximation, in the second
case - to the classical QSS assumption. Both of these approximations allow for hierarchical
realizations, those which include not just two but many relaxation time scales. Such a {\it
multi-scale approach}  essentially simplifies analytical studies of the problem.

\subsection{Thermodynamic criteria for selection of important reactions}

One of the problems addressed by the sensitivity analysis is the selection of the important and
discarding the unimportant reactions. In the paper \cite{BYA77} a simple principle was suggested  to
compare importance of different reactions according to their contribution to the entropy production
(or, which is the same, according to their contribution to $\dot{G}$). Based on this principle,
Dimitrov \cite{Dimitrov82} described domains of parameters in which the reaction of hydrogen
oxidation, ${\rm H}_2+{\rm O}_2+{\rm M}$, proceeds due to different mechanisms. For each elementary
reaction, he has derived the domain inside which the contribution of this reaction is essential
(nonnegligible). Due to its simplicity, this entropy production principle is especially well suited
for analysis of complex problems. In particular, recently, a version of the entropy production
principle was used in the problem of selection of boundary conditions for Grad's moment equations
\cite{Struchtrup98,GKZ02}. For ideal systems (\ref{gfun}), as well, as for the Marcelin-De Donder
kinetics (\ref{HMDD}) the contribution of the $s$th reaction to $\dot{G}$ has a particularly simple
form:
\begin{equation}
\label{dotGs} \dot{G}_{s}=-W_s\ln\left(\frac{W_s^+}{W_s^-}\right),\ \dot{G}=\sum_{s=1}^{r}\dot{G}_s.
\end{equation}

\subsection{Opening}

One of the problems to be focused on when studying closed systems is to prepare extensions of the
result for open or driven by flows systems. External flows are usually taken into account by
additional terms in the kinetic equations (\ref{reaction}):
\begin{equation}
\label{external} \dot{\NN}=V\JJ(\cc)+\bPi(\cc,t).
\end{equation}
It is important to stress here that the vector field $\JJ(\cc)$ in equations (\ref{external})  is the
same, as for the closed system, with thermodynamic restrictions, Lyapunov functions,  etc. The
thermodynamic structures are important for analysis of open systems (\ref{external}), if the external
flow $\bPi$ is  small in some sense, for example, it is a linear function of $\cc$,  has small time
derivatives, etc. There are some general results for such ``weakly open" systems, for example, the
Prigogine minimum entropy production theorem \cite{Prig} and the estimations of possible of steady
states and limit sets for open systems, based on thermodynamic functions and stoihiometric equations
\cite{G1}.

There are general results for another limit case: for very intensive flow the dynamics becomes very
simple again \cite{Yab}. Let the flow has a natural structure:
$\bPi(\cc,t)=v_{in}(t)\cc_{in}(t)-v_{out}(t)\cc(t)$, where $v_{in}$ and $v_{out}$ are the rates of
inflow and outflow, $\cc_{in}(t)$ is the concentration vector for inflow. If $v_{out}$ is
sufficiently big, $v_{out}(t)>v_0$ for some critical value  $v_0$ and all $t>0$, then for the open
system (\ref{external}) the Lyapunov norm exists: for any two solutions $\cc^1(t)$ and $\cc^2(t)$ the
function $\|\cc^1(t) - \cc^2(t)\|$ monotonically decreases in time. Such a critical value  $v_0$
exists for any norm, for example, for usual Euclidian norm $\|\bullet\|^2=(\bullet,\bullet)$.

For arbitrary form of $\bPi$ the system (\ref{external}) can loose all signs of being a thermodynamic
one. Nevertheless, thermodynamic structures often can help in the study of open systems.

The crucial questions are: What happens with slow/fast motion separation after opening? Which slow
invariant manifolds for the closed system can be deformed to the slow invariant manifolds of the open
system? Which slow invariant manifold for the closed system can be used as approximate slow invariant
manifold for the open system? There exists more or less useful technique to seek the answers for
specific systems under consideration \cite{CMIM,ZKD2000}.

The way to study an open system as the result of opening a closed system may be fruitful. In any
case, out of this way we have just a general dynamical system (\ref{external}) and no hints what to
do with it.

\begin{centering}

***

\end{centering}

Basic introductory textbook on physical kinetics of the  Landau  and Lifshitz Course of Theoretical
Physics \cite{LPi} contains many further examples and their applications.

Modern development of kinetics follows the route of specific numerical methods, such as direct
simulations. An opposite tendency is also clearly observed, and the kinetic theory based schemes are
increasingly often used for the development of numerical methods and models in mechanics of continuous
media.

\end{document}